\documentclass[acmsmall]{acmart} %!PN

\def\BibTeX{{\rm B\kern-.05em{\sc i\kern-.025em b}\kern-.08em
    T\kern-.1667em\lower.7ex\hbox{E}\kern-.125emX}}

\title{Mitigate Parasitic Resistance in Resistive Crossbar-based Convolutional Neural Networks}
\acmJournal{JETC}
\acmVolume{SI: Nanoelectronic Device, Circuit, Architecture Design}
\acmArticle{1}
\renewcommand\footnotetextcopyrightpermission[1]{} % removes footnote with conference information in first column
\author{Fan Zhang}
\affiliation{%
  \institution{Binghamton University}
  \city{Binghamton}
  \state{New York}
  \country{USA}}
\email{fzhang27@binghamton.edu}

\author{Miao Hu}
\affiliation{%
  \institution{Binghamton University}
  \city{Binghamton}
  \state{New York}
  \country{USA}}
\email{miaohu@binghamton.edu}

\usepackage{xfrac}  
\usepackage{graphicx}
\usepackage{subcaption}
\usepackage{url}
\graphicspath{ {images/} }
 \newcommand\quotient[2]{
        \mathchoice
            {% \displaystyle
                \text{\raise1ex\hbox{$#1$}\Big/\lower1ex\hbox{$#2$}}%
            }
            {% \textstyle
                #1\,/\,#2
            }
            {% \scriptstyle
                #1\,/\,#2
            }
            {% \scriptscriptstyle  
                #1\,/\,#2
            }
    }

\usepackage{amsmath}
\usepackage{algorithm}
\usepackage{amssymb}
\usepackage[noend]{algpseudocode}
\usepackage{etoolbox}
\usepackage{color}
\usepackage{comment}
\usepackage{citesort}
\usepackage{pifont}% http://ctan.org/pkg/pifont
\newcommand{\cmark}{\ding{51}}%
\newcommand{\xmark}{\ding{55}}%
\newbool{inccomment}
\setbool{inccomment}{true}
\newcommand{\zf}[1]{\ifbool{inccomment}{{\color{blue}#1}}{}}
\newcommand{\hm}[1]{\ifbool{inccomment}{{\color{magenta}#1}}{}}

\newcommand\Tstrut{\rule{0pt}{2.6ex}}       % "top" strut
\newcommand\Bstrut{\rule[-0.9ex]{0pt}{0pt}} % "bottom" strut
\newcommand{\TBstrut}{\Tstrut\Bstrut} % top&bottom struts

\makeatletter
\def\BState{\State\hskip-\ALG@thistlm}
\makeatother
\algnewcommand{\parState}[1]{\State%
  \parbox[t]{\dimexpr\linewidth-\algmargin}{\strut #1\strut}}
  
\begin{document}
\begin{abstract}
Traditional computing hardware often encounters on-chip memory bottleneck on large scale Convolution Neural Networks (CNN) applications.
With its unique in-memory computing feature, resistive crossbar-based computing attracts researchers' attention as a promising solution to the memory bottleneck issue in von Neumann architectures.
However, the parasitic resistances in crossbar deviate its behavior from the ideal weighted summation operation.
In large-scale implementations, the impact of parasitic resistances must be carefully considered and mitigated to ensure circuits' functionality.
In this work, we implemented and simulated CNNs on resistive crossbar circuits with consideration of parasitic resistances. Moreover, we carried out a new mapping scheme for high utilization of crossbar arrays on convolution, and a mitigation algorithm to mitigate parasitic resistances in CNN applications.
The mitigation algorithm considers parasitic resistances as well as data/kernel patterns of each layer to minimize the computing error in crossbar-based convolutions of CNNs.
We demonstrated the proposed methods with implementations of a 4-layer CNN on MNIST, and ResNet(20, 32, and 56) on CIFAR-10. 
Simulation results show the proposed methods well mitigate the parasitic resistances in crossbars. With our methods, modern CNNs on crossbars can preserve ideal(software) level classification accuracy with 6-bit ADCs and DACs implementation.

\end{abstract}




\maketitle

%%%%%%%%%%%%%%%%%%%%%%%%%%%%%%%%%%%%%%%%%%%%%%%%%%%%%%%%%%%%%%%%%%%%%%%%%%%%%%%%

%%%%%%%%%%%%%%%%%%%%%%%%%%%%%%%%%%%%%%%%%%%%%%%%%%%%%%%%%%%%%%%%%%%%%%%%%%%%%%%%
\section{Introduction}

%Deep learning has gained great success in many fields\cite{lecun_deep_2015}. 
Convolutional Neural Networks(CNN) have led to many performance breakthroughs in image classification, video object tracking, and audio processing applications\cite{lecun_deep_2015}. 
Since AlexNet won the ILSVRC 2012 \cite{krizhevsky_imagenet_2012},
%and is more than 10\% lower than other contemporary competitors on top-5 error rate 
CNNs have evolved into many different models, such as GoogLeNet\cite{szegedy_going_2015}, VGG\cite{simonyan_very_2014}, and ResNet\cite{he_deep_2015}.
%, and occupied the first places on ILSVRCs.
Using Graphics Processing Unit(GPU) and specific accelerators to accelerate convolutions play an important role in CNNs' success, since convolutions dominate the overall computation.
% and need hardware with high parallelism on matrix operation.
Interestingly, in many machine learning applications throughput of convolutions is prior than computing accuracy, especially on inference functions \cite{gupta_deep_2015}. 
Inspired by this feature, Nvidia's recent GPUs\cite{ho_exploiting_2017},  Cambricon-X\cite{zhang2016cambricon},  and many other accelerators (such as Cambricon MLU100\cite{CambriconMLU100}, Xilinx FPGA based accelerator\cite{8309067}) start support half-floating point or even 8-bit integer precision to accelerate convolutions without performance degradation in many neural networks.

However, modern computing hardware often encounters on-chip memory bottleneck when dealing with high volume convolutions in large scale CNNs\cite{chandrasekhar_compression_2017,rhu_vdnn:_2016}.
%\hm{Provide reference and evidence for this memeory bottleneck issue}
State-of-art convolution neural networks require tremendous amount of parameters to handle complex tasks. For example, AlexNet has 2.3 million parameters, VGG has 14.7 million parameters, and ResNet-152 has 25.5 million parameters\cite{chandrasekhar_compression_2017}. 
%Such amount of parameters cannot be simply fit into caches.
Storing such amount of parameters into limited caches is impossible.
Therefore, frequent cache flushing and weight loading from off-chip memory (usually DRAM) are usually inevitable, which lead to significant delay and energy cost. 
% Last but not least, DRAM size also the scale of neural networks \cite{rhu_vdnn:_2016}
% Recently, researches unveiled conventional Von Neumann architecture's disadvantage on neuromorphic computing and memory based emerging non-volatile devices' advantage on neuromorphic computing\cite{yu_neuro-inspired_2018}. 

To overcome the memory bottleneck, many researchers show interest in resistive crossbar arrays for the computing-in-memory feature \cite{gao_demonstration_2016,adam_3-d_2017,hu_memristor-based_2018,liu_harmonica:_2016,yu_neuro-inspired_2018,li_analogue_2018}.
The resistive crossbar is defined as a circuit structure with vertical and horizontal metal lines sandwiching a resistive switching material at their intersection. 
The cross-point material could be memristor\cite{strukov2008missing}, phase change memory (PCM) device\cite{wong2010phase}, floating gates\cite{chen1999high}, spintronic device\cite{wolf2001spintronics}, RRAM\cite{wong2012metal}, SRAM\cite{6055294}, or any other devices with programmable resistance.
%Such a crossbar circuit is famous for its unique in-memory computing feature: 
By utilizing Kirchhoff's current law (KCL) and Ohm's law, an ideal resistive crossbar array can carry out analog vector-matrix-multiplication (VMM).
% By collecting analog output current from columns with input voltage signal flowing through the rows, where weights are stored non-volatility as conductance in cross-point devices, large-scale vector matrix multiplication can be done in a single step.
The outputs of analog VMMs are represented as the analog output currents from columns of the crossbar, with input voltage signals flowing through rows, and weights are stored non-volatility as conductance in cross-point. 
In the inference stage, any size of VMMs can be easily done in a single step.
Moreover, since weight storage and weighted multiplication/summation both happen at the same place --- the crossbar array, it enables ultra-high computing efficiency on multiplications between changing vectors(data) and fixed matrices(weight), which is ideal for implementing ultra-low power inference functions of neural networks.   
%In short, crossbar-based architecture can overcome the memory bottleneck in von Neumann architectures on implementing inference functions of neural networks.

However, as array scales up, circuit parasitics deviate crossbar from its ideal linear behavior and bring the non-negligible error to the computing result.
The impact of circuit parasitics, especially parasitic resistances such as wire resistance and interconnect resistance, have been observed and analyzed in many simulations and experiments\cite{hu_dot-product_2016}\cite{ciprut2017modeling}\cite{agarwal2017compensating}.  
Currently, the impact of parasitic capacitance and inductance can be ignored since they mainly affect transient behavior of crossbar, while in-memory computing mainly depends on crossbars' DC behavior.
It is important to consider parasitic resistance in circuit simulations for practical and functional implementations, especially for neural network applications where many large-scale crossbars are used.

In this paper, we investigate crossbar-based large-scale CNN implementation with consideration of parasitic resistances and provide methods to mitigate its impact on convolution accuracy as well as CNN classification accuracy. 
Our major contributions includes: 
\begin{itemize}
\item First, we invent an efficient implementation method to densely map high dimension 4-D kernels to 2-D crossbar arrays. 
Using this method, a crossbar designed for vector-matrix multiplication can be easily adapted for convolution with near-zero hardware overhead.  
%\item Second, we investigate the impact of data and kernels in CNNs on the computing accuracy of memristor-based convolution circuit.
\item Second, we model resistive crossbar array with consideration of parasitic resistances and realistic device models. The simulation result is also verified with experiment data up to  128$\times$64 crossbar size.
\item Third, we study the impact of parasitic resistances and provide a mitigation method to minimize computing error due to parasitic resistances as well as data/kernel patterns in CNNs without  re-training. 
%Our improved conversion algorithm well compensates the impact of circuit parasitic in tested CNNs, and it can be applied to general CNNs since it proves independence of data sparsity and kernel types.
\item Last but not least, we demonstrate our methods for a 4-layer CNN on MNIST, and Resnet-20, 32, and 56 on CIFAR-10. 
%Analog circuit simulation is carefully done at all convolution layers to capture the trends of error propagation through networks.

Comparing to other state-of-the-art crossbar simulators, our work conducts the end-to-end full circuit simulation on modern CNN models and large dataset with the consideration of circuit parasitics and realistic memristor models.
Our work firstly shows that with realistic crossbars and other peripheral circuits (DAC+ADC), how errors will propagate in deep CNNs due to mixed signal processing and nonlinear activations (ReLU).

%The impact of ADC/DAC bit-resolutions on classification accuracy is analyzed. 
Our result shows that, with the proposed implementation and mitigation methods, 8-bit ADC/DAC resolution may be good enough to preserve the software-level classification accuracy in deep CNNs.
%6-bit ADC/DAC resolution is good enough to preserve the software-level classification accuracy on modern CNNs. 
\end{itemize}

The rest of paper is organized as follows: Section II covers the background. Section III details the methodology of the implementation and mitigation methods. Section IV gives simulation result on single crossbar as well as crossbar-based CNNs.  In the end, Section V concludes the paper. 

\section{Preliminary}

\subsection{Convolutional Neural Network}
\begin{figure}[t]
      \centering
     % \framebox{\parbox{3in}
      \includegraphics[width=0.7\textwidth]{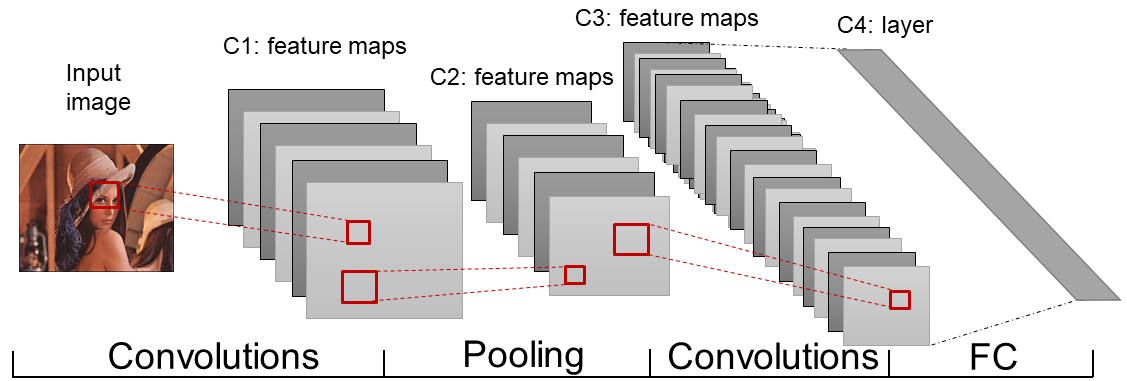}
      \caption{CNN Block Diagram.}
      \label{fig_CNN}
\end{figure}

Fig.\ref{fig_CNN} shows a typical structure of CNN models.
It has three key components: convolutional layers, pooling layers, and fully connected layers\cite{lecun_deep_2015}. 
%Convolutional layer plays the most important role in CNNs. 
Each convolutional layer consists of a set of kernels which have identical sizes. It can be used to detect the local spatial correlation in input patterns. 
The output of convolution is passed to a non-linear activation function, such as ReLU or sigmoid, to form new feature maps. 
Pooling layer acts as a down-sampling filter for the output of the activation function. 
A pooling layer not only reduces the size of the feature map but also merges similar features. 
As a result, it helps to reduce the number of parameters as well as alleviate over-fitting.
The output from pooling layer then feeds into the next convolution layer.
%which helps to reduce the number of parameters need to be trained in the network and to control over-fitting, 
%Usually, pooling layer is also followed with non-linear activation function. 
%After passing the activation function, feature maps are ready for the next convolutional layer or the fully connected layer. 
Fully connected (FC) layers usually appeared at the last few layers of CNNs as the classifier. 
They weight every point in the high-level feature maps and generate the final classification result.

In this work, we are targeting deep residual neural network (ResNet) on resistive crossbar arrays, as it is one of the state-of-the-art CNNs on image classification problems. 
ResNet was firstly introduced in ILSVRC 2015 \cite{he_deep_2015}. 
It keeps the benefit of deeper layers while addressing the degradation issue by optimizing the residual mapping with shortcut connections between layers.
An ensemble of residual nets up to 152 layers has achieved 3.57\% error on the ImageNet test set and won the 1st place on the ILSVRC 2015 classification competition. 
%To evaluate how state-of-the-art CNNs perform on crossbar-based accelerators, ResNet is an ideal test case. 
Fig.\ref{fig_resnet} shows the basic block diagram of ResNet. It combines multiple convolutions, batch normalizations, and rectified linear units(ReLU) together as its basic building block.   
Different from other CNNs, ResNet uses a shortcut to directly add input data to the output result of a block.   
%Right after the shortcut, an adder sums the input and output of the building block and pass the result to the next building block.
Since two data inputs may have different dimensions at the summation stage, a $1 \times 1$ convolution layer is introduced in the shortcut to match the dimensions of two inputs.
The summation result is feed to ReLU and pass to the next Block. 
At the end of ResNet, pooling layer, one or more FC layers, and a softmax layer are used in sequence to generate the final classification result.
By studying and optimizing resistive crossbars for ResNet, we can get more insight into the performance of resistive crossbar-based computing on modern CNNs.
% Due to its high performance/accuracy and simple structure, we adopt a typical ResNet-20 as our research object in this work.
% \begin{figure}[t]
%       \centering
%      % \framebox{\parbox{3in}
%       \includegraphics[width=0.4\textwidth]{images/ResNet.png}
%       \caption{Basic block diagram of ResNet.}
%       \label{fig_resnet}
% \end{figure}

\subsection{Resistive crossbar circuit for VMM computing}
% \begin{figure}[t]
%       \centering
%      % \framebox{\parbox{3in}
%       \includegraphics[width=\columnwidth]{images/xbar_illus.png}
%       \caption{Illustration of resistive crossbar array.}
%       \label{fig_xbar_illus}
% \end{figure}

\begin{figure}
    \centering
\begin{minipage}{0.45\textwidth}
     % \framebox{\parbox{3in}
      \includegraphics[width=0.9\textwidth]{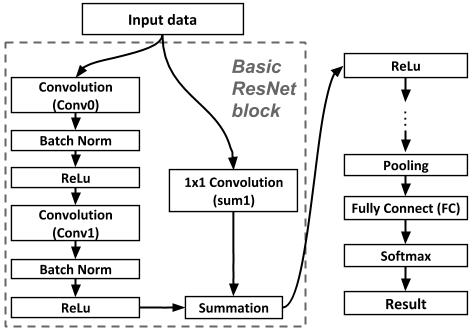}
      \caption{Basic block diagram of ResNet.}
      \label{fig_resnet}
\end{minipage} \hfill
\begin{minipage}{0.5\textwidth}
      \includegraphics[width=\textwidth]{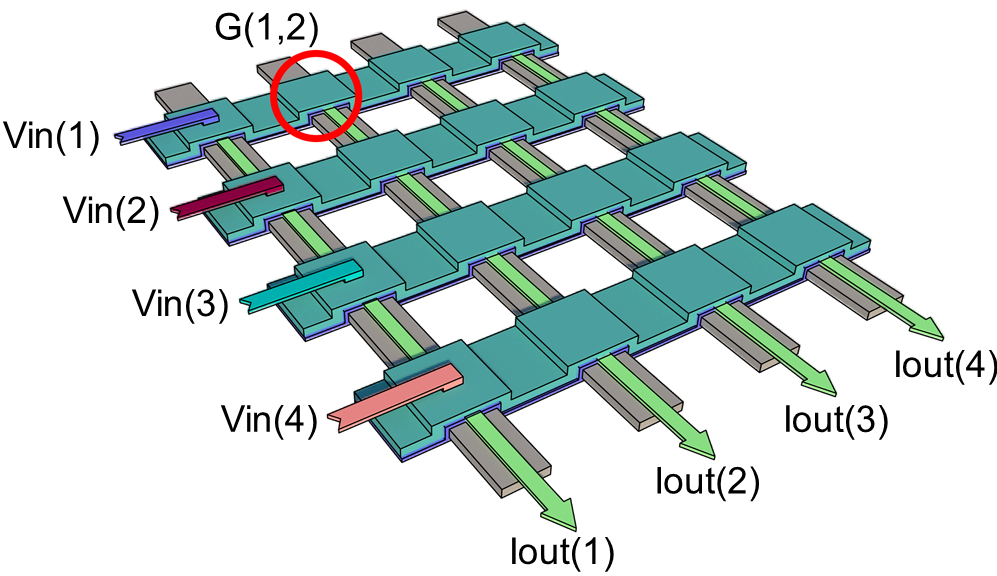}
      \caption{Illustration of resistive crossbar array.}
      \label{fig_xbar_illus}
\end{minipage}
\end{figure}

Fig.\ref{fig_xbar_illus} illustrates the general structure of resistive crossbar array for VMM computing.
In an ideal crossbar, when applying voltage inputs $V$ at the rows simultaneously and read current outputs $I$ from the columns, the input-output relationship of the crossbar can be represented as below:
$$
I = V\textbf{G}
$$
In this way, the analog weighted summation is achieved through Kirchhoff's Current Law and Ohm's Law. 
By mapping input vector $X$ to input voltage $V$, positive matrix \textbf{A} to conductance \textbf{G}, and output current $I$ back to output result $Y$, a memristor crossbar can be regarded as an analog VMM module to realize $Y=X\textbf{A}$ in one step. 
Note that $I = V\textbf{G}$ is only valid for ideal crossbar where parasitic resistances can be ignored, and device conductance is independent of voltage/current. 

In realistic crossbar arrays, circuit parasitics, including wire resistance, input/output resistance, device I-V non-linearity, and parasitic capacitance, deviate crossbar's behavior from ideal vector-matrix multiplication. 
For instance, the wire resistance degrades signal along the row and column wire, so the device on the further corner will receive signal less than expected due to increased wire resistance. 
Input/Output resistance act similarly as wire resistance, but they could cause even larger signal degradation since they are usually caused by pass transistors and more resistive than wires. Meanwhile, cross-point device I-V non-linearity affects its multiplication accuracy, as its conductance is no longer independent of voltage or current. 
The impact of capacitance and inductance can be ignored at current stage since they mainly affect the transient behavior of crossbar while we are using crossbar's DC behavior to realize analog VMM at relatively low frequency (10M to 100MHz).

\subsection{Mapping algorithms for crossbar array}
% \hm{Need a more compete review on mapping algorithms}
 
When using the crossbar array as the VMM circuit, the first problem is how to map the matrix onto the crossbar. Since the intersection conductance and column output current cannot be the negative value, the mapping method needs to able to address the negative issue. 
One common way is using the positive and negative power supply, and matrices are mapped on crossbar array in absolute form. For the negative element in the matrix, the negative input voltage is applied to the corresponding position.
%Mapping algorithms for crossbar array map matrix values to conductance of memristor crossbars. 
Recently, Chris Yakopcic  et.al have proposed a way to implement convolution and CNN on memeristor-based crossbar \cite{7727302}\cite{728081ex-suit}.
In their ex-suit process, convolutional kernel $\mathbf{G}$ has to be divided into two part, positive $\mathbf{G^{+}}$ and negative $\mathbf{G^{-}}$. 
All negative values in $\mathbf{G^{+}}$ have been replaced by zero.
% and all positive values in $G-$ have been replaced by zero. 
And those negative values have been mapped in $\mathbf{G^{-}}$ as positive conductance.
By providing $\mathbf{G^{-}}$ the identical inputs to $\mathbf{G^{+}}$ but with reversed sign, memristor-based crossbar can deal with the matrix which contains negative values.

\cite{7479502} also maps the absolute value of the matrix on crossbar array. However, it needs two adjacent columns of the crossbar to represent a single column of that matrix. 
One represents the positive part, and another represents the negative part. Then the two columns have output currents $I^{+}$ and $I^{-}$ respectively. Final output = $I^{+} - I^{-}$.

\cite{ji2019fpsa} uses another way to solve this negative issue in their ReRAM crossbar based PE(processing element). 
The positive and negative part of the matrix are still mapped on two adjacent columns of a crossbar. 
But the negative voltage input is unnecessary in this design. Instead, an extra subtracter component is needed for every pair of positive and negative columns. 
Another change in this design is that it uses spiking schema to improve the output precision. $2^{n}$ spikes are used to represent a number of n bits.

In the above mapping schemes, either extra overheads or redundant crossbar areas are needed.
Moreover, for all above implements, a rigorous full circuit simulation with consideration of parasitic resistances on a large enough CNN is still missing.

\subsection{Crossbar based Neural Network simulator}
There exist a lot of crossbar-based simulators for neural network applications with the consideration of parasitic-effect, stuck-at-fault, thermal noise and more. 
However, most of those simulators are not experiment-verified and even over-simplified for the non-ideal effects.
Table \ref{tab_NN_simulator} compares state-of-the-art crossbar-based NN simulators. 
Our simulator is the only circuit simulator that experimentally verified and works on large-scale dataset and NN models.

MNSIM\cite{xia_mnsim:_2016} and NeuroSim\cite{chen_neurosim:_2018} are two famous ReRAM crossbar simulators.
They count many non-ideal effects into consideration but they do not focus on accurate crossbar computation output due to lack of full analog circuit simulation.
However, they could give a reasonable power/area estimation for crossbar based design.
Moreover, they are designed for general crossbar-based neural networks not optimized for convolution neural networks. 

PytorX\cite{He:2019:NIA:3316781.3317870} and FTNNA\cite{Liu:2019:FNN:3316781.3317742} are designed for neural network applications with the consideration of non-ideal effects.
They also have the ability to alleviate the impact of those non-ideal effects and have been demonstrated on large datasets.
However, such rescue ability is given by the error tolerance of NN models rather than the optimization of simulators, since both need to re-training the model to assuage the non-ideal effects. 
In practical we cannot guarantee that all crossbars have same non-ideal effects and those effects may vary because of different crossbar size, stored conductance matrix, wire resistance, and etc.
Re-training for every crossbar device is a plausible way to solve such non-ideal effects, but they may not always practical due to time/energy limitation. 
Moreover, online training for large-scale NNs requires deep understanding and very accurate modeling of memristor dynamic behaviors, which is still far from enough at this moment. 
Even we do training with existing memristor models, the learning we can get from it is questionable because the model is still very different from realistic device behaviors and does not account the impact of realistic process variations and noises.

In this work, we developed an experiment-verified simulator to conduct the end-to-end SPICE-level analog crossbar circuit simulation with the consideration of circuit parasitics and realistic memristor models.
Conversion and calibration methods are used to mitigate the parasitic effects in crossbar-based computing. 
Comparing to other simulators, our work not only shows the impact of  non-ideal effects on NN model accuracy, but also explains the reasons behind such accuracy degradation by investigating the error propagation in crossbar-based deep CNNs.
In this way, our work provides a more solid upper-bound for the inference performance of crossbar-based CNNs respecting different design parameters.

\begin{table}[tb]
\caption{Crossbar based NN simulator comparison}
\label{tab_NN_simulator}
\resizebox{\textwidth}{!}{
\begin{tabular}{|l|c|c|c|c|l|l|l|}
\hline
Simulator     & \multicolumn{1}{l|}{\begin{tabular}[c]{@{}l@{}}Experiment\\ -verified\end{tabular}} & \multicolumn{1}{l|}{\begin{tabular}[c]{@{}l@{}}Full circuit \\ simulation\end{tabular}} & \multicolumn{1}{l|}{MNIST} & \multicolumn{1}{l|}{CIFAR-10} & NN type & Non-ideal effects                                                                                                                              & Method to rescue IR-Drop                                                                \\ \hline
MNSIM\cite{xia_mnsim:_2016}         & \xmark                                                                     & \xmark                                                                         & \xmark                   & \xmark              & MLP,CNN & \begin{tabular}[c]{@{}l@{}}Consider wire resistance \\ into simplified parallel \\ resistance\end{tabular}                                     & \multicolumn{1}{c|}{\xmark}                                                         \\ \hline
NeuroSim\cite{chen_neurosim:_2018}      & \xmark                                                                     & \xmark                                                                         & \cmark                 & \xmark              & MLP     & \begin{tabular}[c]{@{}l@{}}Wire RC considered in \\ nearby synaptic cell\end{tabular}                                                          & \begin{tabular}[c]{@{}l@{}}Setting a small  \\ IR\_DROP\_TOL\end{tabular}               \\ \hline
PytorX\cite{He:2019:NIA:3316781.3317870}        & \xmark                                                                     & \xmark                                                                         & \xmark            & \cmark                   & CNN     & \begin{tabular}[c]{@{}l@{}}Simplified nodal analysis \\ for wire resistance, \\ stuck-at-fault, \\ thermal \& shot noise\end{tabular}          & \begin{tabular}[c]{@{}l@{}}Re-training with \\ IR-drop consideration\end{tabular}       \\ \hline
FTNNA\cite{Liu:2019:FNN:3316781.3317742}         & \xmark                                                                     & \xmark                                                                         & \cmark                 & \cmark                   & CNN     & \begin{tabular}[c]{@{}l@{}}Programming error, \\ stuck-at-fault\end{tabular}                                                                   & \begin{tabular}[c]{@{}l@{}}Retraining with error \\ correction output code\end{tabular} \\ \hline
DPE\cite{hu_dot-product_2016}           & \cmark                                                                          & \cmark                                                                              & \cmark                 & \xmark              & MLP     & \begin{tabular}[c]{@{}l@{}}SPICE-level analysis with \\ transistor models, circuit \\ parasitics, and actual \\ memristor models.\end{tabular} & Conversion                                                                              \\ \hline
Our simulator & \cmark                                                                          & \cmark                                                                              & \cmark                 & \cmark                   & CNN & \begin{tabular}[c]{@{}l@{}}SPICE-level analysis with \\ transistor models, circuit \\ parasitics, and actual \\ memristor models.\end{tabular} & \begin{tabular}[c]{@{}l@{}}Conversion \& \\ Calibration\end{tabular}                    \\ \hline
\end{tabular}
}
\end{table}

\section{\bf Methodology}

There are two challenges ahead of using resistive crossbars for convolution operations. 
The first challenge is how to transform high dimensional convolution kernels to 2-D matrices for resistive crossbars. 
The most straightforward way converts convolution kernels to Toeplitz matrices. However, the Toeplitz matrix is sparse, and its sparsity significantly grows with larger kernels.
Directly mapping this sparse matrix to resistive crossbars, will cause large hardware overhead as well as computing accuracy concerns.
The second challenge is how to mitigate parasitic resistances in crossbar-based computing.
As introduced before, parasitic resistances in crossbar will cause nonlinear signal degradation at each cross-point devices, lack of considering its impact will cause unacceptable errors in computing result of large crossbar arrays.
To solve both challenges, we present the dense mapping as the new mapping method to fully unitize crossbar arrays with near zero hardware overhead, and the mitigation algorithm to compensate parasitic resistance in large-scale crossbars.

\subsection{Dense mapping for crossbar-based convolution}

The critical step of mapping an arbitrary matrix $A$ on crossbar is how to deal with negative values in \textbf{A}, as conductance cannot be negative. 
We first shift all elements in \textbf{A} by a constant $A_{Shift}$ to make sure no negative value in A. 
Such a shift can be easily removed at the final step by subtracting $A_{Shift} \times sum(X)$. 
Where $sum(X)$ is the summation of input. By doing so, we do not need to part matrix into positive and negative sub-matrices, and any VMM can be performed on the crossbar with much less overhead.

When we use crossbar for VMM computing in digital circuit environment, ADC/DAC are necessary to convert input data to analog voltage and output current to digital form.
This process usually assumes to be 10ns to 100ns, which is usually limited by the ADC speed. 
To get sum(X), an accumulator works in parallel with the DAC->crossbar->ADC system with similar speed. 
For small or medium size crossbars, digital accumulator would be more efficient to calculate the summation of input vectors; For large crossbars, an analog accumulator can be implemented by using an additional column in the crossbar array where its devices are programmed to the same conductance, and it is sensed by an additional ADC. 
With either method of accumulator design, the cost and hardware overhead on calculating sum(X) is much lower than using two crossbars or two devices to track negative values. 

In CNN, each convolution layer has a 4-D kernel. For example, a 3*3*3*16 kernel means it has 16 sets of 3*3*3 3-D kernels, and each 3-D kernels contains three 2-D kernels for the three channels (RGB) of color images. 
To perform convolution on memristor crossbar, we need to convert high dimensional convolution into 2-D VMM. 
It is well known that 2D convolution can be implemented using matrix multiplication by converting kernel to a Toeplitz matrix. 
%Since kernels are fixed after training, It is better to convert kernels to Toeplitz matrices rather than data. 
Toeplitz matrices are sparse matrices which contain many zeros, so we named this method as \textit{sparse mapping}. 

\begin{figure}[t]
      \centering
      \includegraphics[width=0.5\textwidth]{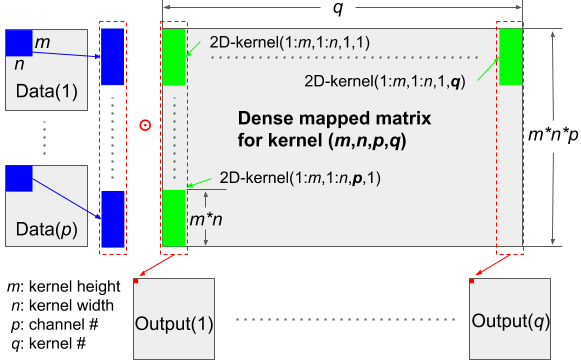}
      \caption{Dense mapping for input data size (j,k,p), where j,k,p means data height, width, and channel, respectively. The circuit needs j*k iterations to process the input data.}
      \label{Fig_densemap}
\end{figure}
   
However, sparse mapping has three major issues:
First, memristor crossbar is by nature a dense array structure with positive values,  not efficient for sparse matrix implementation.
Mapping a sparse matrix to crossbar means that those memristors assigned with zero would do nothing but adding error to the final result due to the leakage current, as well as waste circuit area.    
Second, sparse mapping requires a huge crossbar array, which is not always feasible, and vulnerable to noise and defects.  
%However,We can partition one large matrix into several small size matrix, it introduces another noise due to multiple crossbars are needed to perform single image(each crossbar's accuracy may vary).
%Partition and implement may be a solution, but it also means that we need to combine partitioned result from different crossbar arrays, and error/noise accumulation would be another concern.
Third, sparse mapping requires a sizeable peripheral circuit to support the enormous array. 
Since the peripheral circuit dominates the total power/area, the hardware implementation of sparse mapping would be too costly to afford.

In contrast to sparse mapping, we developed a new method, named as \textit{dense mapping}, targeting on using small and dense memristor crossbars to implement convolutions efficiently.
Fig.\ref{Fig_densemap} illustrates the concept of dense mapping.
%The dense matrix mapping method is much more efficient on space aspect and easy to tune output accuracy than the sparse way.
Each 2-D kernel is unrolled to a vector and mapped to one column of crossbar so that one crossbar can implement multiple 2-D kernels as long as it has enough columns. 
For input signal, only data within the convolution window are fed to the row inputs of crossbar arrays. 
%For data with multiple channels, 2-D kernels for different channels are stacked into the same column, as input from different channels can be supplied to different rows and weighted summed together.
When input data has multiple feature channels, each output feature channel needs a 3-D convolution kernel. A 3-D convolution kernel can be treated as many 2-D kernels. 
Every 2-D kernel corresponds to one input channel. The output of 3-D convolution is given by performing convolution on all 2-D input channels and add them together. Thus, we can unroll every 2-D kernel in the 3-D kernel as the same order. 
It is then cascading all vectors together to form a single column on a crossbar. 
An input shift register stores the unrolled input data within convolution window(multiple input channels cascaded like the kernel on a crossbar) at the current iteration to feed to the crossbar, then updating its storage as the window moves through the entire data space.
The convolution results for data within the convolution window are collected at the column outputs of the crossbar.
In this way, a single convolution kernel with one stride on both horizontal and vertical direction needs $(j - m + 1)$*$(k - n 
+ 1)$ iterations where $j$, $k$ / $m$, $n$ are data/kernel height and width respectively. 
Since multiple input channels have been compressed in a single column on a crossbar, the input channel number doesn't impact the iteration time.

Comparing dense mapping to sparse mapping, it is a trade-off between time complexity and space complexity.
Sparse mapping uses much more extra hardware to produce the result in parallel without iteration.
From the data movement aspect, the least movement mapping method is sparse mapping. It uses extra space in crossbar to not only store weight in there but also perform the kernel window movement.
However, its efficiency exponentially drops as data/kernel size scales up because more devices in a rectangular crossbar are unused for increasing data/kernel size. 

And much larger crossbar and more ADC/DACs are required for sparse mapping a large data/kernel. 
Table \ref{tab_dense_sparse} compares the overhead of dense\&sparse mapping on common used data/kernel sizes from ResNet.
In Table \ref{tab_dense_sparse}, crossbar, ADC/DAC parameters are adopted from ISAAC\cite{7551379}. We exponentially scaled the DAC to 8-bit, then proportionally scaled the area and power for crossbar, ADC, and DAC respectively.
Although we could partition huge matrix into multiple small matrices\cite{Lin:2019:LSR:3287624.3287715}, sparse mapping still needs ~100x numbers of DACs\&ADCs than dense mapping. 

\begin{table*}[]
\caption{Overhead comparison for dense\&sparse mapping}
\label{tab_dense_sparse}
\resizebox{\textwidth}{!}{
\begin{tabular}{|l|l|l|l|l|l|l|l|}
\hline
Kernel Size & \begin{tabular}[c]{@{}l@{}}xbar size\\ (dense\&sparse)\end{tabular} & \begin{tabular}[c]{@{}l@{}}xbar power\\ (mW)\end{tabular} & \begin{tabular}[c]{@{}l@{}}xbar area\\ ($mm^{2}$)\end{tabular} & \begin{tabular}[c]{@{}l@{}}ADC \\ number\end{tabular} & \begin{tabular}[c]{@{}l@{}}DAC \\ number\end{tabular} & \begin{tabular}[c]{@{}l@{}}Total power\\ (mW)\end{tabular} & \begin{tabular}[c]{@{}l@{}}Total area\\ ($mm^{2}$)\end{tabular} \\ \hline
3x3x3x16    & 27 x 16                                                             & 0.06                                                      & 0.000005                                                 & 1                                                     & 27                                                    & 29.06                                                      & 0.001                                                     \\ \hline
            & 3072 x 14400                                                        & 6480                                                      & 0.54                                                     & 113                                                   & 3072                                                  & 9778                                                       & 0.806                                                     \\ \hline
3x3x16x16   & 144 x 16                                                            & 0.34                                                      & 0.000028                                                 & 1                                                     & 144                                                   & 146.34                                                     & 0.007                                                     \\ \hline
            & 16384 x 14400                                                       & 34560                                                     & 2.88                                                     & 113                                                   & 16384                                                 & 51170                                                      & 3.712                                                     \\ \hline
3x3x32x32   & 288 x 32                                                            & 1.35                                                      & 0.000113                                                 & 1                                                     & 288                                                   & 291.35                                                     & 0.014                                                     \\ \hline
            & 8192 x 6272                                                         & 7526.4                                                    & 0.6272                                                   & 49                                                    & 8192                                                  & 15816.4                                                    & 1.034                                                     \\ \hline
3x3x64x64   & 576 x 64                                                            & 5.4                                                       & 0.00045                                                  & 1                                                     & 576                                                   & 583.4                                                      & 0.026                                                     \\ \hline
            & 4096 x 2304                                                         & 1382.4                                                    & 0.1152                                                   & 18                                                    & 4096                                                  & 5514.4                                                     & 0.311                                                     \\ \hline
\end{tabular}
}
\end{table*}

%But it needs large size of crossbar which may sacrifice result accuracy. 

\begin{figure}
      \centering
      \includegraphics[width=\textwidth]{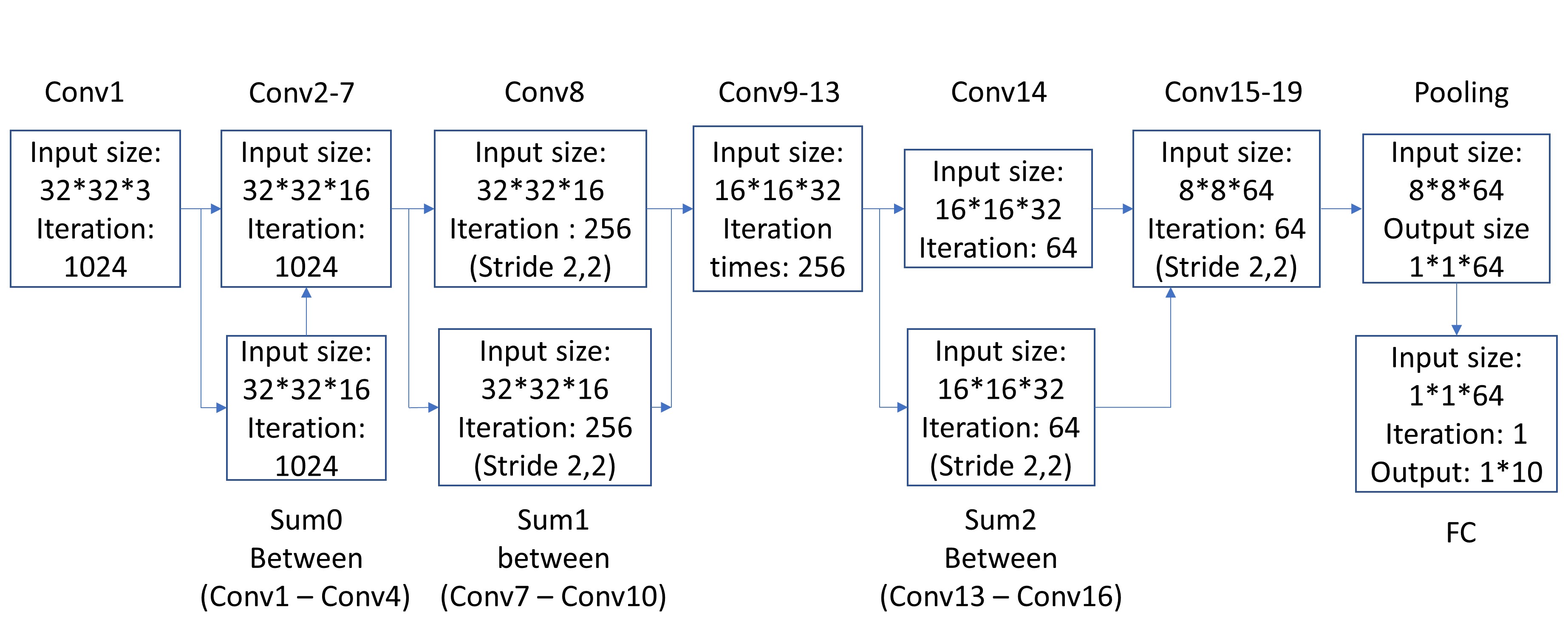}
      \caption{Flowchart of a 32 by 32 image in ResNet20}
      \label{fig_res20iters}
   \end{figure}

Therefore, dense mapping is an adequate and more practical method comparing to sparse mapping. 
It not only achieves 100\% usage of devices but also easy to implement and provide sufficient performance in speed for CNN applications. 
From Fig.\ref{fig_res20iters}, one classification inference in ResNet-20 needs 9089 iterations in sequential, if no parallel copies of hardware are used. 
Note that summation (sum\#) is in parallel of convolutions, so it's not counted in total iterations.
Assuming crossbar runs at 100 MHz \cite{hu_dot-product_2016}, for each classification the convolution part takes only 0.09 ms, which is fast enough for real-time classification requirement.

\subsection{Crossbar simulation with parasitic resistances}

\begin{figure}
    \centering
\begin{minipage}{0.5\textwidth}
      \centering
     % \framebox{\parbox{3in}
      \includegraphics[width=\textwidth]{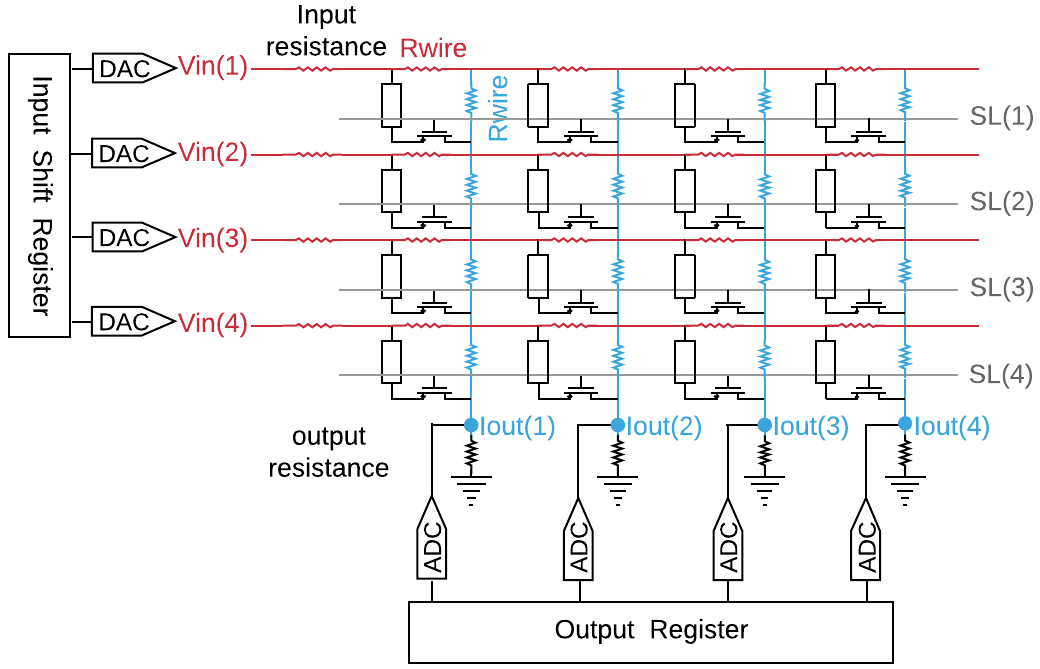}
      \caption{Crossbar model with consideration of parasitic resistance.
      }
      \label{fig_xbar_model}
\end{minipage} \hfill
\begin{minipage}{0.45\textwidth}
      \centering
     % \framebox{\parbox{3in}
      \includegraphics[width=0.9\textwidth]{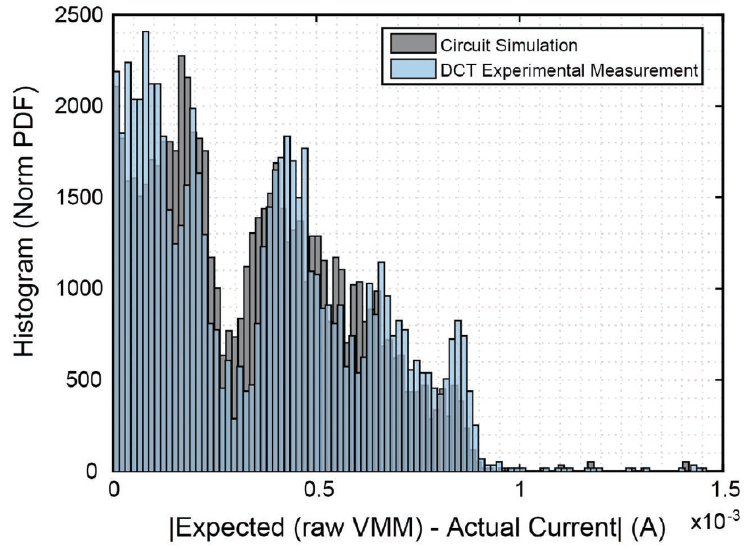}
      \caption{1T1M simulation and experiment result \protect\cite{hu_memristor-based_2018}. Wire segment resistance is calibrated to 1$\Omega$, transistor and memristor models are calibrated with in-array device test.}
      \label{fig_1T1Msim}
   \end{minipage}
   \end{figure}

Fig.\ref{fig_xbar_model} shows the circuit structure of  one-transistor-one-memristor (1T1M) crossbar for vector matrix multiplication (VMM).
In this structure, memristor crossbar is the core component as it enables analog current weighted summation, which leads to ultra-efficient VMM operation. 
A $m \times n$ memristor crossbar is made by $m$ row and $n$ column metal wires where memristors formed at intersecting points. 
Each memristor can be tuned to arbitrary conductance within its programmable range.
To enable precise, non-disturbing tuning in large crossbar arrays, 1T1M cell 
%(Fig.\ref{fig_xbar}(b).) 
is necessary to program the entire array with arbitrary conductance matrix \textbf{G}.

Besides memristor crossbars, DACs/ADCs are also essential to ensure accurate VMM operation. First, they are necessary to integrate memristor-based analog VMM module into the digital environment, as functions like pooling, ReLU, and normalization still need digital implementation; 
%\zf{Second, the mapping between physical behavior $I=VG$ and mathematical operation $Y=XA$ could to be configured at DAC and ADC stage; }
Second, the calibration step can be performed at the ADC/DAC stage to improve crossbar result further.
Last but not least, ADC provides quantization, which is helpful on filtering out the output error of memristor-based analog VMM modules, to prevent error accumulation inter- and intra-layers in CNNs.

\begin{figure}[t]
      \centering
     % \framebox{\parbox{3in}
      \includegraphics[width=0.7\textwidth]{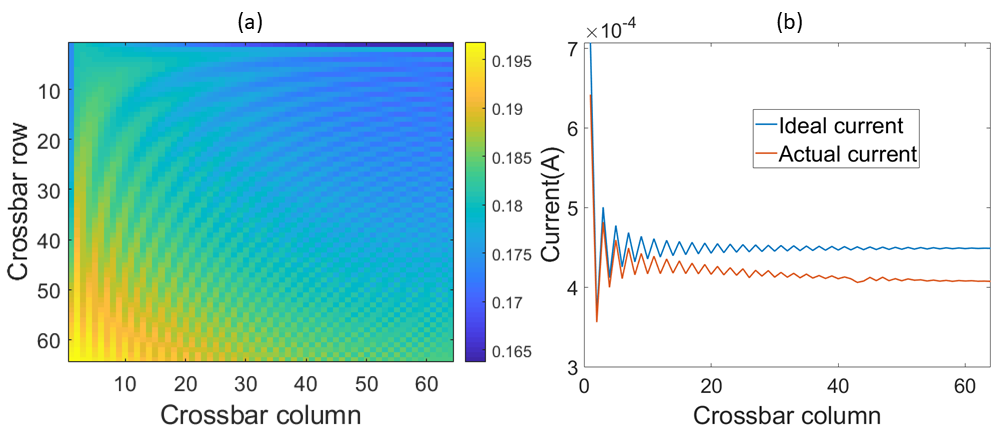}
      \caption{Signal degradation along rows and columns in crossbar. a). 0.2V signal imported into crossbar from the very left point on each row. b). Current degradation along columns in crossbar. Due to parasitic resistance, further column has lower voltage and current than ideal.
      }
      \label{fig_signal_degradation}
\end{figure}

% Fig.\ref{fig_xbar_model} shows the resistive crossbar modeling with consideration of parasitic resistance. Comparing to previous simulation work on memristor crossbar arrays, we modeled parasitic resistance on wires and interconnects.  
Comparing to previous simulation work on memristor crossbar arrays, we considered parasitic resistances including wire resistances, input/output resistances, and intersection resistances in our simulation model.
With consideration of parasitic resistances, we can observe the signal degradation along rows and columns in the array, as shown in Fig.\ref{fig_signal_degradation}. 
The low input signals applied on further corner memristors also make the column current output lower than expectation.
To guarantee the actual output close to the ideal output, we adopt an experiment verified method\cite{hu_memristor-based_2018}. 
Instead of mapping conductance $\mathbf{G}$ to a crossbar, we map a tweaked conductance matrix $\mathbf{G'}$ with consideration of parasitic resistances in the crossbar.
To find $\mathbf{G'}$ for a $m \times n$ crossbar, we first need to formulate all $3m \cdot n+2m+2n$ KCL equations from the crossbar circuit model to solve all node voltages, including $m \cdot n$ cross-point top nodes, cross-point middle nodes (between memristor and access transistor), cross-point bottom nodes, and $2m + 2n$ boundary nodes on both end of horizontal lines and vertical lines. 
Then as conductance $\mathbf{G'}$ has $m \cdot n$ unknown variables, we add $m \cdot n$ new equations $I_{m} = V \odot \mathbf{G}$ as new limitations to force each cross-point to pass  the ideal current. 
Where $I_{m}$ is the cross-point current for each memristor, $V$ is the ideal cross-point voltage between top nodes and bottom nodes, and $\odot$ denotes Hadamard product, or say element-wise multiplication.  
Finally with $4m \cdot n+2m +2n$ nonlinear equations, $\mathbf{G'}$ can be solved with HSPICE or any nonlinear equation solver.
As shown in Fig.\ref{fig_1T1Msim}, the simulation result well matches the experiment result of a 128$\times$64 array.

% \begin{figure}[t]
%       \centering
%      % \framebox{\parbox{3in}
%       \includegraphics[width=0.3\textwidth]{images/1T1Mxbar_VMM.png}
%       \caption{1T1M crossbar for VMM operation. Each black square is an 1T1M cell. 
%       %(c) shows the ramping ADC design.
%       }
%       \label{fig_xbar}
% \end{figure}

% \begin{figure}[t]
%       \centering
%      % \framebox{\parbox{3in}
%       \includegraphics[width=0.4\textwidth]{images/1T1Msim.png}
%       \caption{1T1M simulation and experiment result \protect\cite{hu_memristor-based_2018}. Wire segment resistance is calibrated to 1$\Omega$, transistor and memristor models are calibrated with in-array device test.}
%       \label{fig_1T1Msim}
%   \end{figure}

%\input{Crossbar_Simulator}
\subsection{Mitigation algorithm for parasitic resistance}

Algorithm \ref{alg_1} summarizes the flow of crossbar-based convolution with the mitigation algorithm. 
Table \ref{Table_functions} explains the important functions in the algorithm. 
After initialization, if the kernel is already mapped and converted onto crossbars, it will directly jump to the computing step to simulate crossbar-based convolution. 
So we only need to mapping conductance matrix once for CNN inference.   

\begin{algorithm}
\caption{Mitigation Algorithm}\label{euclid}
\begin{algorithmic}[1]
\State Initialization: setup crossbar parameters(wire resistance, memristor conductance range, etc...)
\State Get a batch of input data
\State If kernel $K$ is mapped and converted, jump to {\bf computing}.
%\State Begin \textbf{mapping} 
\State Dense mapping kernel $K$ to conductance matrix \textbf{G}
%\State End \textbf{mapping}
%\State Begin \textbf{conversion}
\State Optimize conversion signal
\State $InputVectors \gets \text{Partition input data}$
\State $CaliSample \gets \text{Random pick from } InputVectors$
\State $\mathbf{G'} \gets \text{Conversion}(\textbf{G},V_{conv})$
\State $P \gets \text{GetCaliPara}(\mathbf{G'},CaliSample)$
%\State End \textbf{conversion}
\State Begin \textbf{computing}
\While {$\sim$end of $InputVectors$}
\State $V_{in}[i] \gets InputVector[i]$
\State $I_{out}[i] = \text{CrossbarSim}(\mathbf{G'}, V_{in}[i])$
\State $Output[i] = \text{Calibration}(I_{out}[i],InputVector[i],P)$
\State $i \gets i + 1$
\EndWhile
\State $\text{Convolution result} \gets \text{reshape}(Output)$
\State End \textbf{computing} 
\end{algorithmic}
\label{alg_1}
\end{algorithm}

\begin{table}[t]
\caption{Explanations for functions in algorithm \ref{alg_1}}
\centering
\begin{tabular}{|l|p{10.5cm}|}
\hline
\textbf{Function} & \textbf{Explanation}\TBstrut\\ \hline
CrossbarSim       & Experiment verified crossbar simulator from \cite{hu_memristor-based_2018}\TBstrut\\ \hline
Conversion      & Solve $V_{conv} \cdot \textbf{G} = \mathrm{CrossbarSim}(V_{conv}, \textbf{G'})$ to get \textbf{G'}.\TBstrut \\ \hline
GetCaliPara      & Get 1st order poly fitting result $P$ by fitting crossbar\Tstrut \  output to ideal output of calibration samples.\Bstrut\\ \hline
Calibration      & Use $P$ and $InputVector[i]$ to map $I_{out}[i]$ to\Tstrut \  $Output[i]$. Here $InputVector[i]$ is needed, because in VMM $Y = X\textbf{A}$, if \textbf{A} contains negative values, $Y$ can be calculated by $Y = X(\textbf{A} + c) - c*\mathrm{sum}(X)$, while c is a large enough scalar to shift \textbf{A} to all positive.\Bstrut\\ \hline
\end{tabular}
\label{Table_functions}
\end{table}

\begin{figure}[t]
      \centering
     % \framebox{\parbox{3in}
      \includegraphics[width=0.7\textwidth]{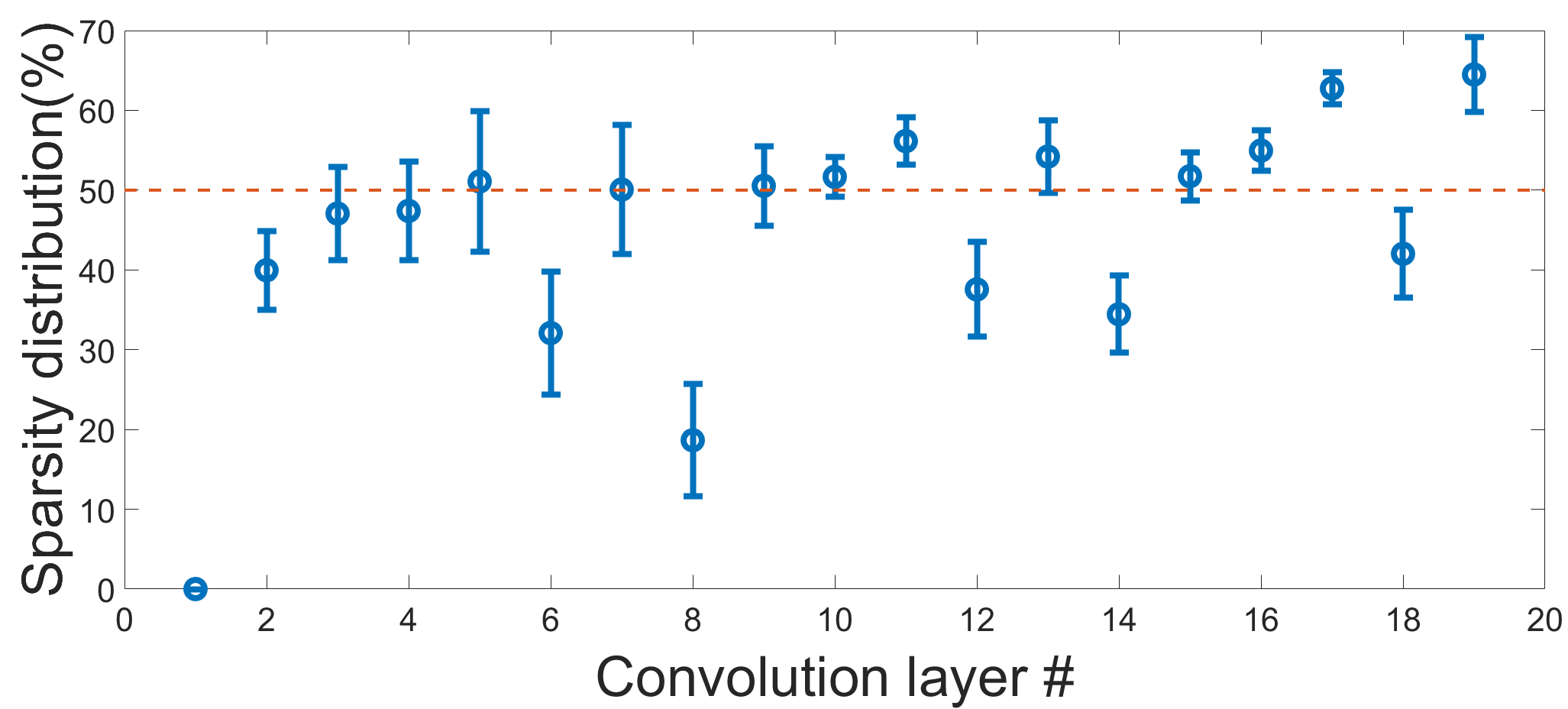}
      \caption{Input data sparsity at each convolution layer of ResNet-20}
      \label{fig_data_sparsity}
   \end{figure}
   
\begin{figure}[t]
      \centering
     % \framebox{\parbox{3in}
      \includegraphics[width=0.7\textwidth]{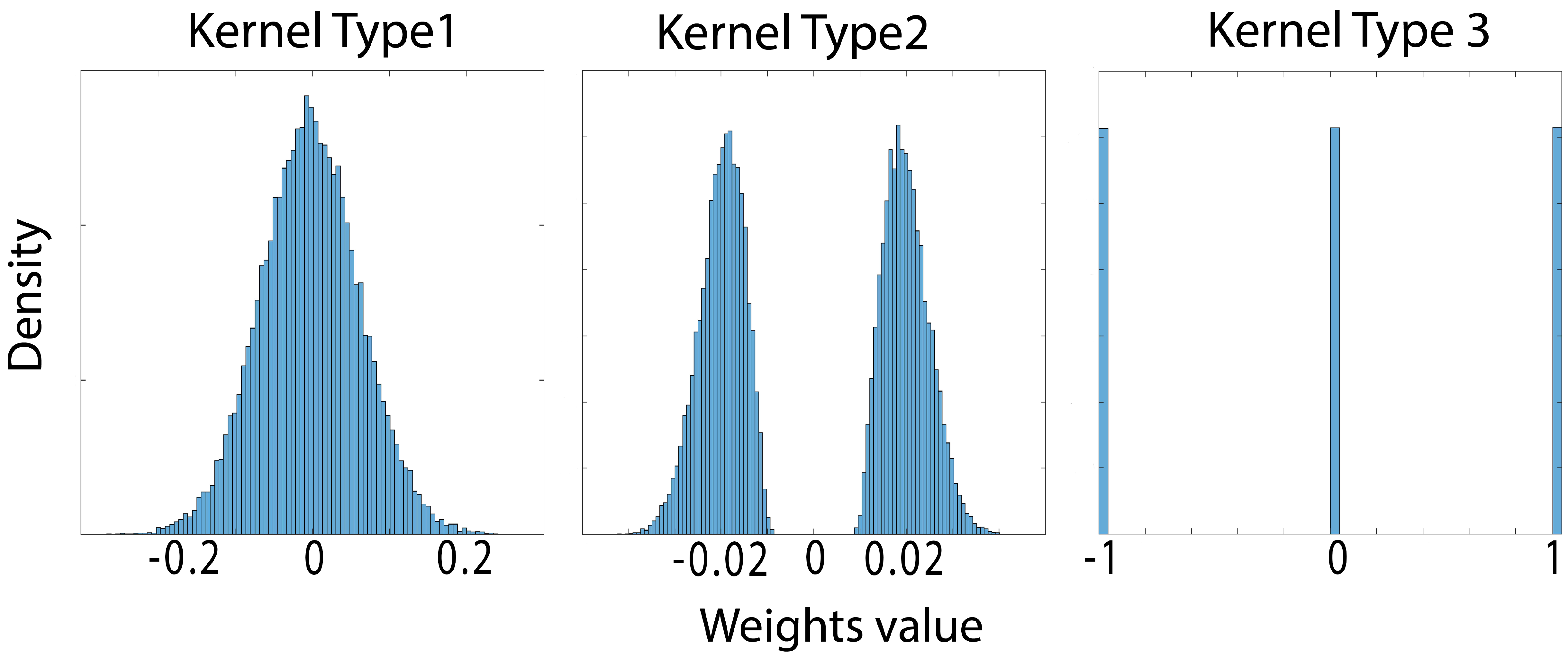}
      \caption{Three kernel types in CNN with different limitations on weight}
      \label{fig_kernel_type}
    \end{figure}
   
\subsubsection{\bf Data and kernel pattern}
%Data in ResNet has high sparsity due to ReLU. 
Input data has high sparsity after the ReLU layer.
Fig. \ref{fig_data_sparsity} shows the data sparsity at each convolution layer of ResNet-20. 
The impact of data sparsity should be considered when choosing the conversion signal as well as gathering calibration samples. 

Similarly, we found that kernels in CNN have different distributions.
In Fig.\ref{fig_kernel_type} we list three typical kernel types regarding to their weight value distributions.
%Usually there are three typical kernel types:
Kernel type 1 refers to a weight distribution close to Gaussian. Usually, it happens when training algorithms put no particular limitation on weight values, such as ResNet. 
Kernel type 2 refers to the training algorithm that preventing weight values goes near zero\cite{han_deep_2015}. 
Kernel type 3 refers to Ternary Neural Networks where weights can only be -1, 0(sparse), or 1\cite{li2016ternary}. 
It's worth investigating how different kennel types in CNN impact the quality/precision of crossbar-based convolution.   
%Here we limits our discussion to the three typical ones due to page limitation.

\subsubsection{\bf Optimize conversion signal}
To better quantify the computing accuracy, we define Output Error and Relative Error as below:
$$
\mathrm{Output~Error} = \frac{\mathrm{ActualOutput} - \mathrm{IdealOutput}}{\mathrm{Output~Range}} 
$$
$$
\mathrm{Relative~Error} = \mathrm{abs(Output~Error)}
$$
% $$
% \mathrm{Relative~Error} = \mathrm{Output~Error}/\mathrm{Output~Range} 
% $$
While ActualOutput is the crossbar output in our simulator. IdealOutput is the expected correct result of the same kernel and input data.
%Absolute Error is the absolute error of crossbar output, 
Output Range is the ideal convolution output range for each kernel.    
Relative error is the absolute value of output error, and it can be converted to output bit accuracy as below: 
$$
\mathrm{Bit~Accuracy} = log_2(1/\mathrm{Relative~Error} + 1)
$$
The conversion step is fine-tuning crossbar conductance from \textbf{G} to \textbf{G'} to compensate the parasitic resistances.
The original conversion algorithm\cite{hu_dot-product_2016} takes the maximum input vector as its conversion signal, which works well with dense matrix and dense input signals. 
However, in CNN, we need to consider the sparsity of data/kernel to optimize the conversion signal. 
By testing different conversion signals across different kernels, we notice that the amplitude of conversion signal is critical, while the sparsity of conversion signal is not as important.
Fig.\ref{fig_diffamp} shows the relative error distribution with different conversion signal amplitudes in crossbar with size $144 \times 16$. 
We found that a conversion signal with too large amplitude (all 1) will cause overcompensation for crossbar conductance matrix due to circuit parasitic, and a too-small conversion signal (all 0.001) do not have enough compensation and both of them result in obvious output error. 
A bad conversion signal may cause hundreds of times error than a good conversion signal($\pm 20\%$ in all 1 signal versus $\pm0.2\%$ in all 0.1 signal).
So for crossbar size $144 \times 16$, all 0.1 signal appears to be the best conversion signal, and other conversion signals close to it generate similar error distribution.

\subsubsection{\bf Calibration}
\begin{figure}
    \centering
    \begin{minipage}{0.48\textwidth}
      \centering
     % \framebox{\parbox{3in}
      \includegraphics[width=\textwidth]{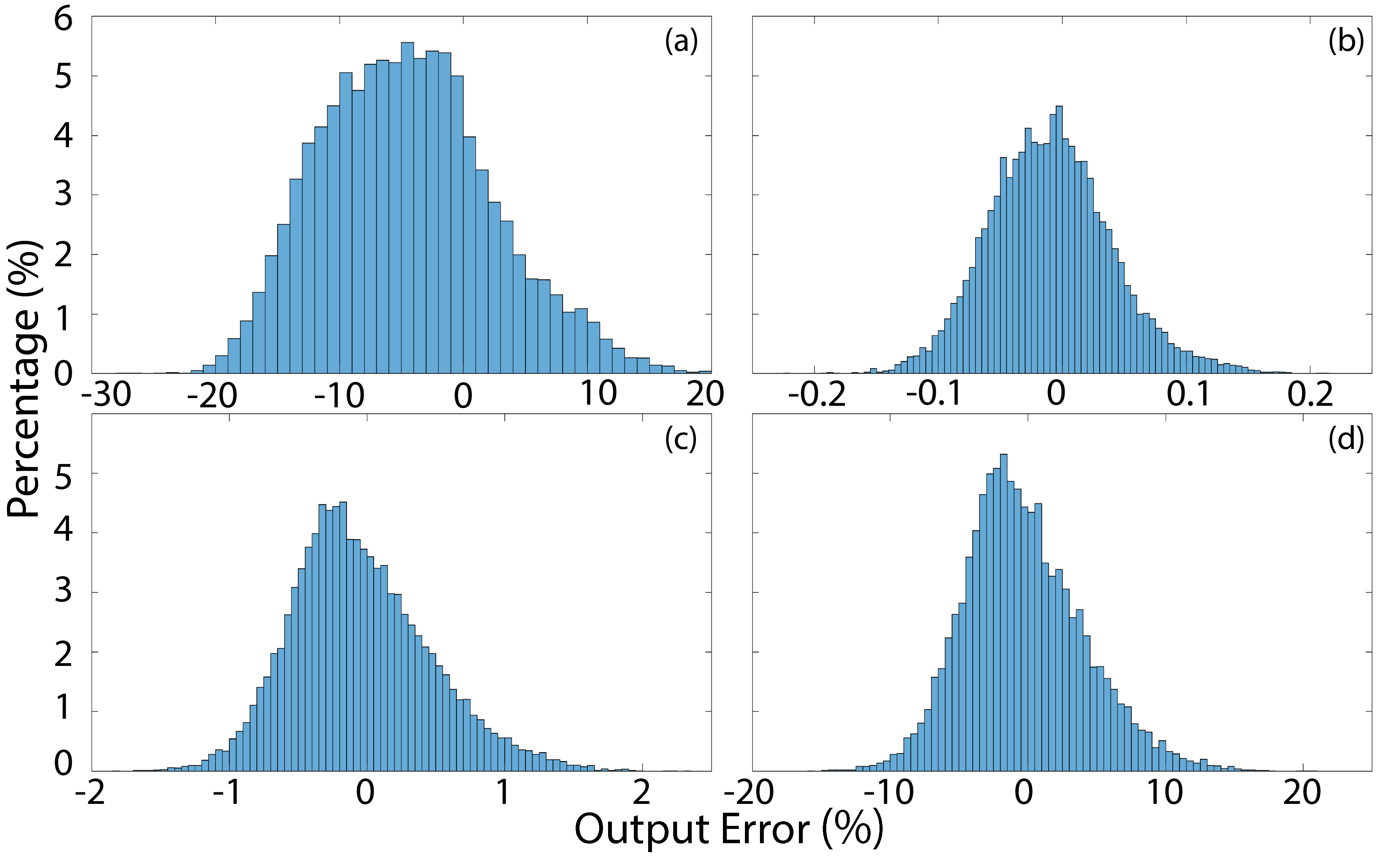}
      \caption{Output error with different conversion signal amplitudes on crossbar(size: $144 \times 16$), (a),(b),(c),(d) are conversion signals with all 1, 0.1, 0.01, 0.001 respectively.}
      \label{fig_diffamp}
   \end{minipage}\hfill
\begin{minipage}{0.48\textwidth}
      \centering
     % \framebox{\parbox{3in}
      \includegraphics[width=\textwidth]{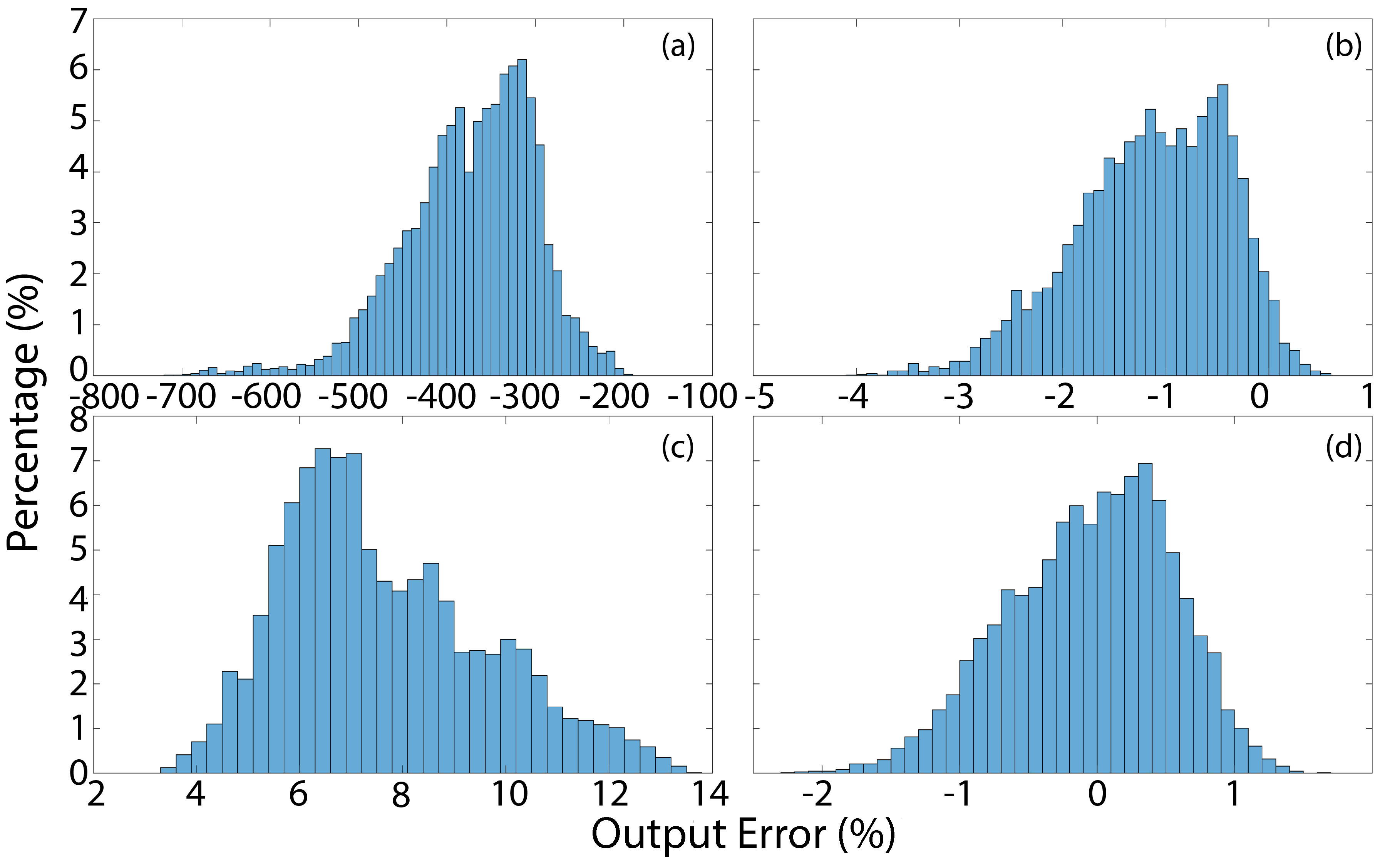}
      \caption{Calibration with different signal types on crossbar(size: $288 \times 32$), (a):Non-calibration, (b):random calibration signal, (c):random but decreased value(rand/cali\_iters),(d):chosen from shuffled input.}
      \label{fig_diff_signal_type}
   \end{minipage}
\end{figure}

In addition to the original conversion algorithm, we add a calibration stage to improve the result further. 
It randomly picks ten samples from the input data set and runs a 1st order polynomial fit to fit crossbar output to ideal output. 
The generated fitting vector $P$ is fixed per crossbar and can be easily embedded in ADC/DAC configurations.
Fig.\ref{fig_diff_signal_type} shows the relative error with different calibration signals, and shuffled input patterns achieve the best result in all four sets of different calibration signals.

\section{\bf Result}

\subsection{ Simulation setup}

In this work, convolution and fully-connected(FC) layers are implemented by analog crossbars with the digital interface. 
Other functions, such as pooling, ReLU, batch normalization, etc., are processed by digital circuits. 
The CNN is offline-trained, then its kernels are converted to the conductance of crossbar for inference.
Similar to our previous work\cite{zhang2018memristor}, our crossbar parameters are listed below: lowest resistance $R_{on}$ = 15k$\Omega$, highest resistance $R_{off}$ = 300k$\Omega$, wire resistance per segment is set to 1$\Omega$, input/output resistance of crossbar are set to 1$\Omega$. Input voltage range is [0, 0.4V]. 
The sensing voltage for device conductance is 0.2V.   
%The crossbar simulation tool is adapted from \cite{hu_memristor-based_2018}.
The CNN framework used in this work is MatConvNet\cite{vedaldi2015matconvnet}.

\subsection{ Individual convolution layer simulation}

We first run circuit simulation at individual convolution layer to study the impact of input data sparsity, kernel type and crossbar size on convolution accuracy. 

\begin{figure}[tp]
      \centering
     % \framebox{\parbox{3in}
      \includegraphics[width=0.7\textwidth]{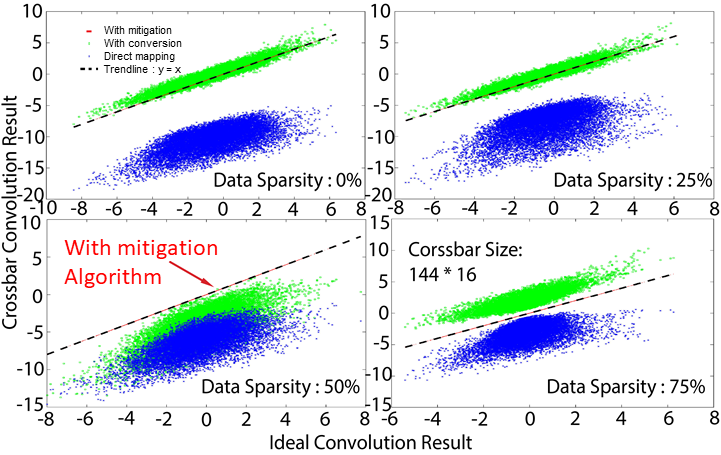}
      \caption{Comparing three methods with different signal sparsities for crossbar size $144 \times 16$}
      \label{fig_comparsion}
   \end{figure}

Fig. \ref{fig_comparsion} shows the output distribution with different input sparsity and different mapping algorithms in a $144 \times 16$ crossbar, which stores a 3*3*16*16 convolution kernel.
If just using linear mapping without any consideration of circuit parasitics, crossbar result deviates far always from the ideal outcome.
Our mitigation algorithm demonstrates better performance compared to the original conversion algorithm in all cases. The result overlaps well on the $y = x$ trend-line with different input sparsity.  
The original conversion algorithm can force the output to get close to the ideal output. But as the sparsity goes up, the original conversion algorithm loses its ability to amend the result.

\begin{figure}
        \centering
    \begin{minipage}{0.48\textwidth}
      \centering
      \includegraphics[width=\textwidth]{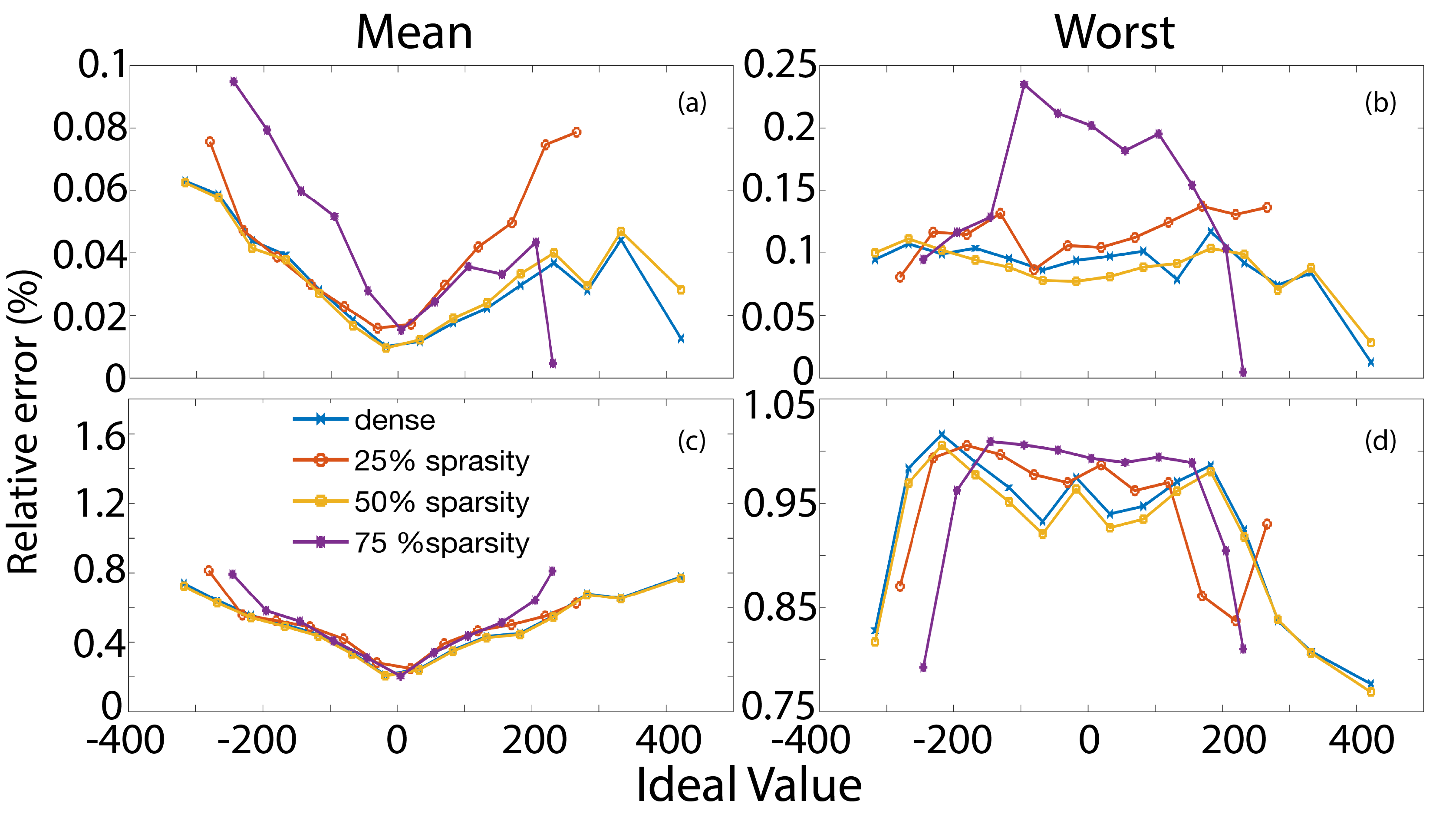}
      \caption{Relative error of $27 \times 16$ crossbar. (a),(b): mitigation algorithm, (c),(d): original conversion algorithm}
      \label{fig_27x16}
    \end{minipage}\hfill
    \begin{minipage}{0.48\textwidth}
       \centering
      \includegraphics[width=\textwidth]{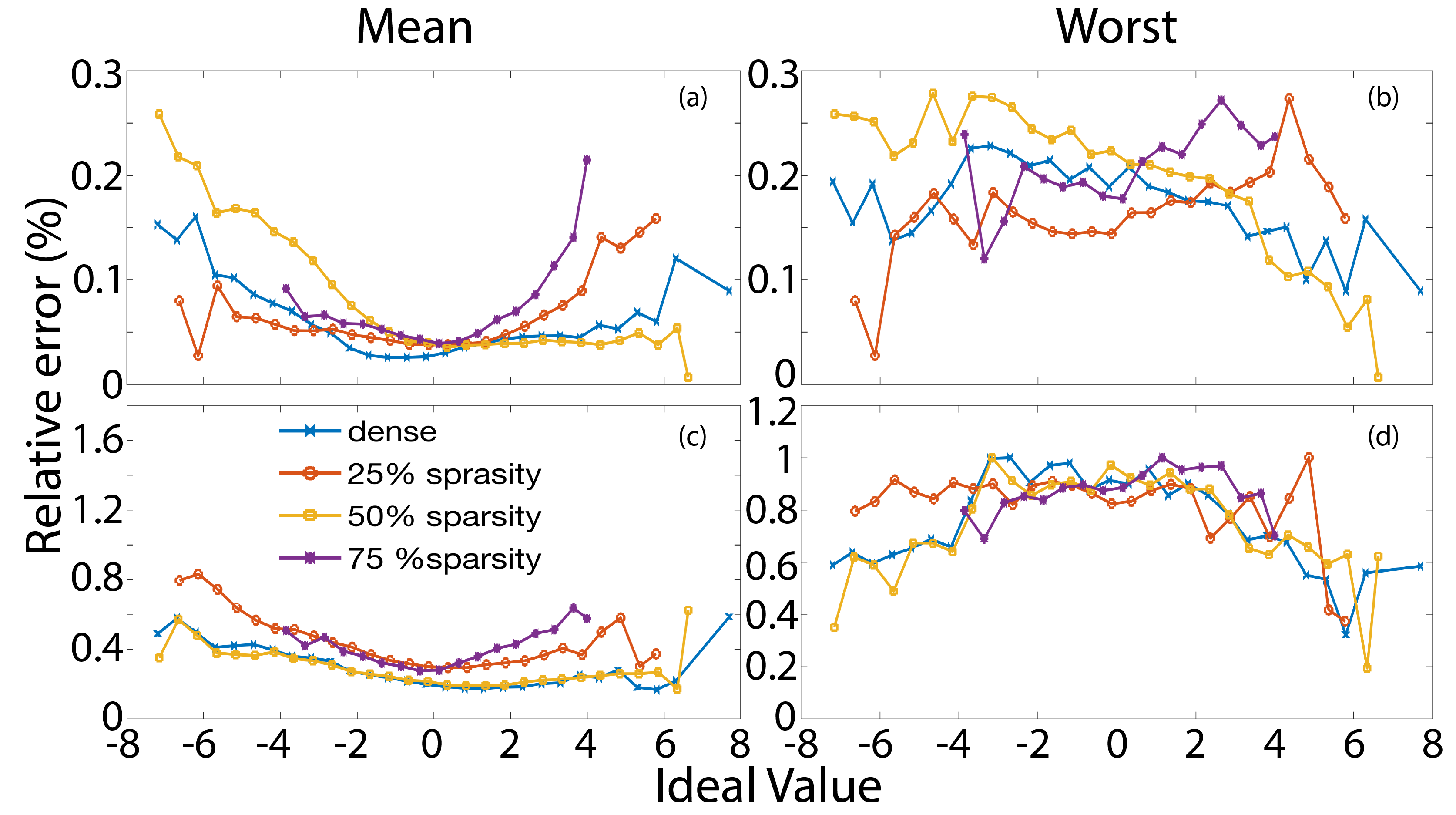}
       \caption{Relative error of $144 \times 16$ crossbar. (a),(b): mitigation algorithm,
       (c),(d): original conversion algorithm}
       \label{fig_144x16}
       \end{minipage} 
    \end{figure}

\begin{figure}
    \centering
       \begin{minipage}{0.48\textwidth}
      \centering
      \includegraphics[width=\textwidth]{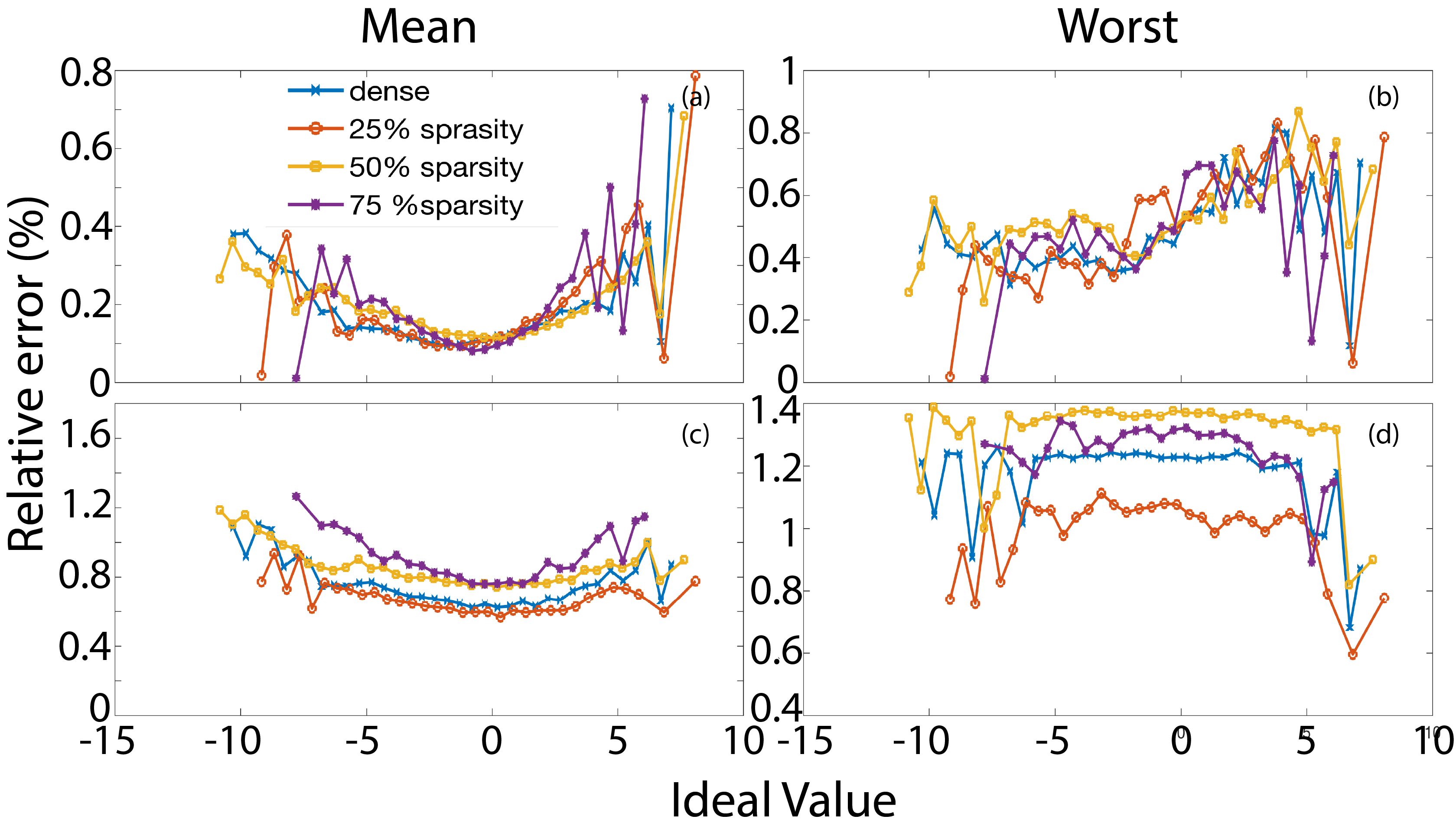}
      \caption{Relative error of $288 \times 32$ crossbar. (a),(b): mitigation algorithm, (c),(d): original conversion algorithm}
      \label{fig_288x32}
   \end{minipage}\hfill
  \begin{minipage}{0.48\textwidth}
       \centering
       \includegraphics[width=\textwidth]{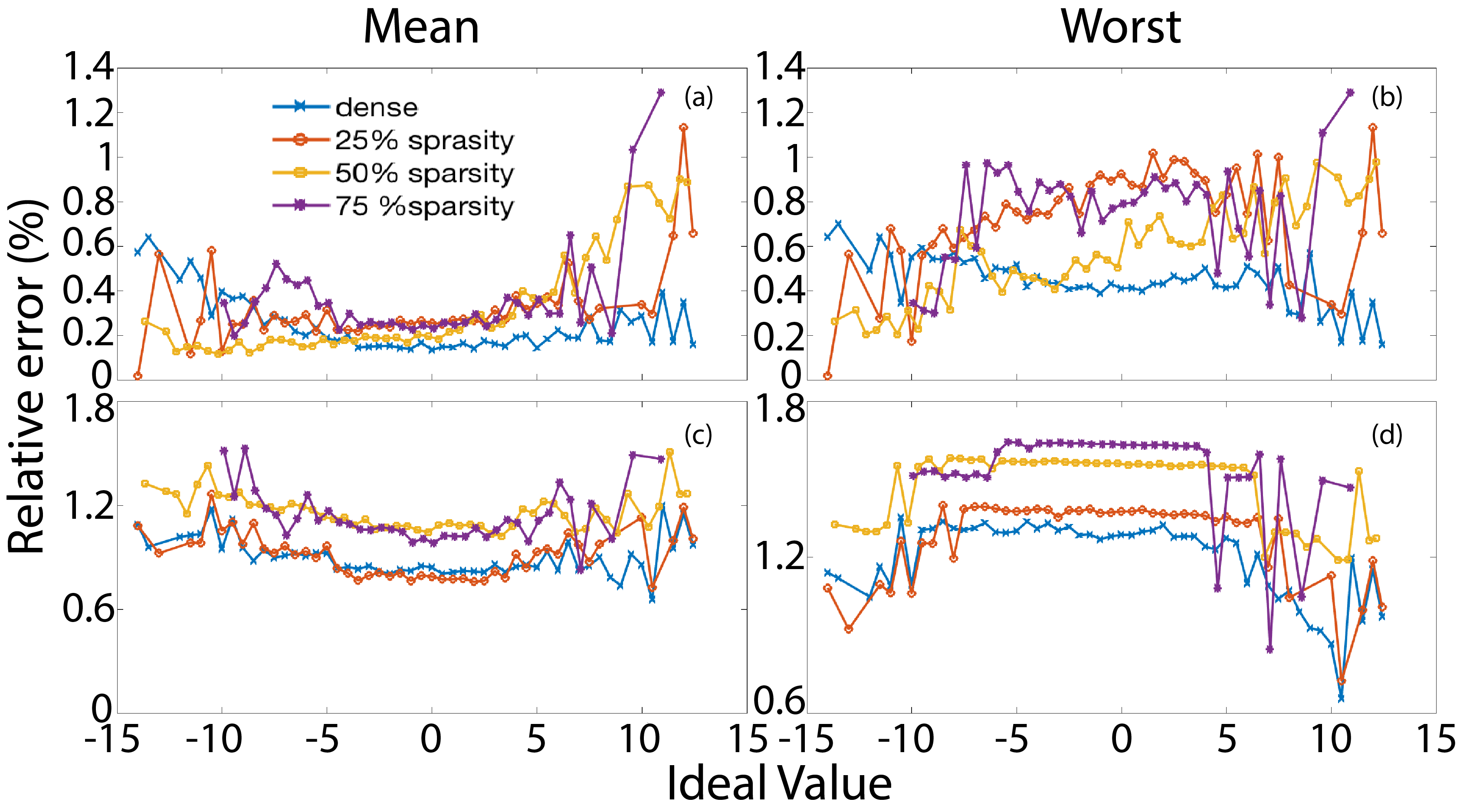}
       \caption{Relative error of $576 \times 64$ crossbar. (a),(b): mitigation algorithm, (c),(d): original conversion algorithm}
       \label{fig_576x64}
   \end{minipage}
\end{figure}

\begin{figure}[t]
      \centering
     % \framebox{\parbox{3in}
      \includegraphics[width=0.7\textwidth]{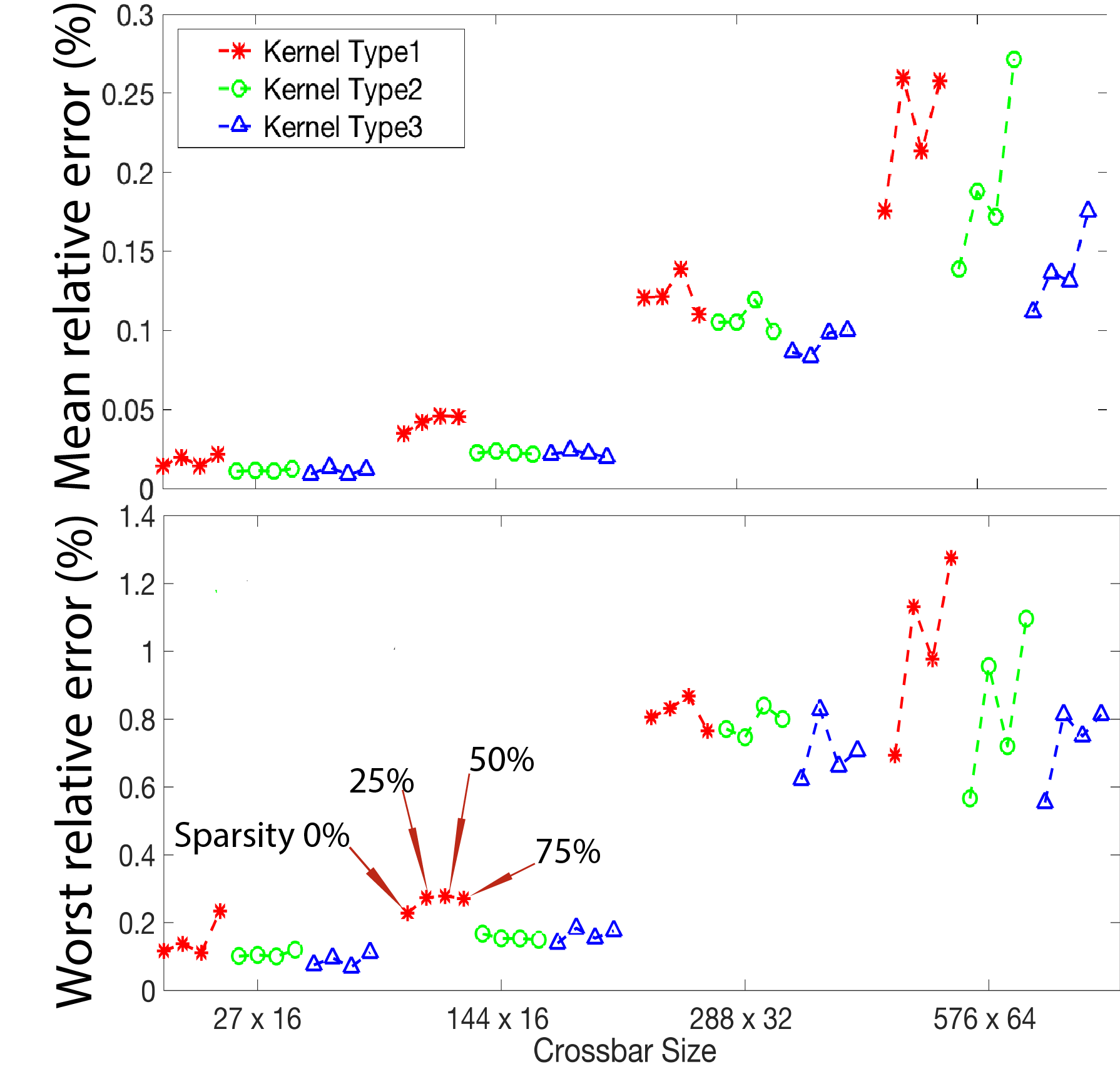}
      \caption{Mean/Worst relative error with different kernel types, data sparsities and crossbar sizes}
      \label{fig_kernel type}
   \end{figure}

Fig.\ref{fig_27x16}, \ref{fig_144x16}, \ref{fig_288x32}, \ref{fig_576x64} illustrate the impact of input data sparsity on different size of crossbar. 
There are three observations: first, our method provides $\sim$50\% better overall accuracy than the original conversion algorithm. 
Second, our method gives a lower mean relative error when the ideal value is small. 
Third, our method minimizes the impact of data sparsity comparing to the original conversion algorithm.
Fig.\ref{fig_kernel type} summarizes the mean/worst relative error across the aforementioned three kernel types with different input sparsity and crossbar sizes. In short, the result shows that our method is independent of kernel types and data sparsity, and achieves the best accuracy overall. 
% \begin{figure}[t]
%       \centering
%      % \framebox{\parbox{3in}
%       \includegraphics[width=0.5\textwidth]{images/conversion_2.png}
%       \caption{Relative error with different conversion signal amplitudes on crossbar(size: $144 \times 16$), (a),(b),(c),(d) are conversion signals with all 1, 0.1, 0.01, 0.001 respectively.}
%       \label{fig_diffamp}
%   \end{figure}\hfill
% \begin{figure}[t]
%       \centering
%      % \framebox{\parbox{3in}
%       \includegraphics[width=0.5\textwidth]{images/cali_relative2.png}
%       \caption{Calibration with different signal types on crossbar(size: $288 \times 32$), (a):Non-calibration, (b):random calibration signal, (c):random but decreased value(rand/cali\_iters),(d):chosen from shuffled input.}
%       \label{fig_diff_signal_type}
%   \end{figure}

    % \begin{figure}[tp]
    %   \centering
    %  % \framebox{\parbox{3in}
    %   \includegraphics[width=0.5\textwidth]{images/27.png}
    %   \caption{Relative error of $27 \times 16$ crossbar. (a),(b): mitigation algorithm, (c),(d): original conversion algorithm}
    %   \label{fig_27x16}
    % \end{figure}
   
    % \begin{figure}[tp]
    %   \centering
    %   % \framebox{\parbox{3in}
    %   \includegraphics[width=0.5\textwidth]{images/144.png}
    %   \caption{Relative error of $144 \times 16$ crossbar. (a),(b): mitigation algorithm,
    %   (c),(d): original conversion algorithm}
    %   \label{fig_144x16}
    %   \end{figure}  
    
%   \begin{figure}[tp]
%       \centering
%      % \framebox{\parbox{3in}
%       \includegraphics[width=0.5\textwidth]{images/288.png}
%       \caption{Relative error of $288 \times 32$ crossbar. (a),(b): mitigation algorithm, (c),(d): original conversion algorithm}
%       \label{fig_288x32}
%   \end{figure}

%   \begin{figure}[tp]
%       \centering
%       \includegraphics[width=0.5\textwidth]{images/576.png}
%       \caption{Relative error of $576 \times 64$ crossbar. (a),(b): mitigation algorithm, (c),(d): original conversion algorithm}
%       \label{fig_576x64}
%   \end{figure}

\subsection{ End-to-end CNN circuit simulation}         

CNN contains more than one layer, any error in one layer will propagate and accumulate to the next layer. 
It is essential to analyze how error propagates and accumulates in deep neural networks, and evaluate their impact on the final classification result.
For MNIST\cite{726791}, We trained a CNN with four convolution layers. 
Table.\ref{simpleNN_size} summarizes the kernel information, corresponding crossbar sizes, and the conversion signal for each layer.
Note that for larger crossbars, conversion signal with smaller amplitude tends to ease the calculation of $\mathbf{G'}$, and provides $\mathbf{G'}$ with better quality (higher VMM accuracy).
It is because, in the conversion process, conversion signals are used for the initialization of node states in crossbar simulation. For example, we initialize all top voltage nodes to be its row input voltages.
Large conversion signal is okay for small to medium size crossbar arrays since the signal degradation along wires is not too significant. 
However, in large crossbar arrays, the node voltage of devices in further corners are significantly different from the ideal values due to notable parasitic resistance. 
Initializing these nodes with ideal values not only makes the solver takes more iteration to complete, but also may lead to non-convergence issues due to lousy initialization, observed as extremely large or even negative values in $\mathbf{G'}$. 
In practice, we found that as the crossbar size goes up, the lower conversion signal helps the algorithm to compute the $\mathbf{G'}$ quickly in the appropriate range. 

\begin{figure}[t]
      \centering
     % \framebox{\parbox{3in}
      \includegraphics[width=0.8\textwidth]{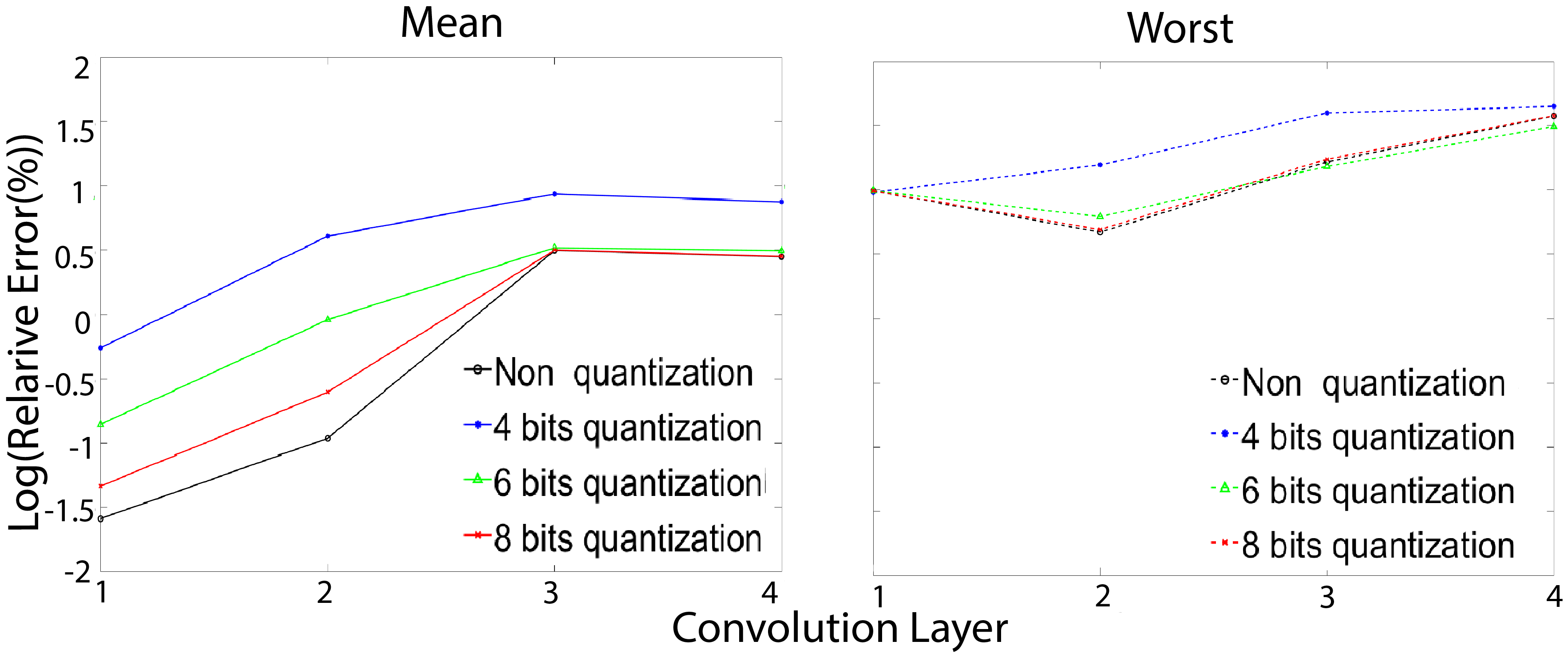}
      \caption{Error propagation along layers in a simple neural network(4 convolution layers) with MNIST dataset, solid lines represent mean error, dash lines represent worst error.}
      \label{fig_mnist_error}
   \end{figure}

Fig.\ref{fig_mnist_error} shows the error propagation at each convolution layer for MNIST. 
%Fig.\ref{fig_ResNet_result} shows the classification result.
Different ADC/DAC quantization bit-resolutions are applied to restrict error propagation. 
As a result, we can see that 6 and 8-bit quantization both can prevent error accumulation after the third layer. 
Another observation is that Non-quantization (direct analog forwarding) makes the error even lower than 8-bit. 

We further tested 4000 images from MNIST validation dataset with different quantization setting. Testing result is given in Table \ref{more_result}.
With 8 bit quantization, final classification accuracy remaining at 98.8\%, which is less than 1\% difference than the ideal (software) result (99.1\%).

We further implement the state-of-art CNN, ResNet on CIFAR-10\cite{Krizhevsky09} to explore the impact of error propagation in modern deep neural networks.
% \hm{Didn't we also introduced ResNet before?}\zf{Aforementioned ResNet is in general, here i am focus on the kernel size of 20,32,56.}
ResNet-20, 32, and 56 all consist by following types of convolution layers: first a convolution layer connected with input has convolution kernel size 3*3*3*16 then a bunch of layers followed with kernel size 3*3*16*16, 3*3*32*32, and 3*3*64*64 respectively. 
Second, additional convolution layer is added between different kernel size to match the matrix dimension. 
Its size depends on nearby convolution layers but with fixed convolution window(1 by 1) on feature map. 
The last convolution layer is the fully connected layer which has the kernel size 1*1*64*10. Besides the last FC layer and dimension match layer, convolution layers in ResNet use the same 3*3 convolution window in their every channel. 
It means for each channel, only nine memristors in the same column is needed to perform the convolution operation. 
Due to such small convolution kernel and few feature channels in ResNet, even in 56 layers, ResNet the largest crossbar needed is still the same in ResNet-20 (576$\times$64).

In table\ref{resnet_size}, we give the convolution kernel sizes, corresponding crossbar sizes and the calibration signal amplitude for ResNets. 
In ResNet, some bypass branches use 1*1 convolution to match the matrix dimension. 
For those dimension match layer, since kernels are tiny, we use the 0.1 calibration signal for all cases. 
Although crossbar size in ResNet is much smaller than CNN we used with MNIST dataset, extra layers make it needs more crossbars for inference. Circuit simulation for ResNet is slower than the CNN mentioned above for MNIST. 

Limited by long circuit simulation time (Each test image takes about 5 minutes), We use a subset of CIFAR-10 which contains 150 images as our test set. 
In Table \ref{more_result}, our experiments show that 8-bit quantization is a good balance between error suppression and information preservation, as it achieves even slightly better classification result than software.
4-bit quantization causes the error accumulates through all layers and leads to a significant drop in final classification accuracy.

   \begin{table}[]
   \caption{Simple CNN convolution layer kernel size, crossbar size , and corresponding conversion signal amplitude }
\begin{tabular}{|l|l|l|l|}
\hline
Layer & Kernel Size & Crossbar Size & Conversion signal \\ \hline
Conv1 & 5*5*1*20    & 25 * 20       & 0.1              \\ \hline
Conv2 & 5*5*20*50   & 500 * 50      & 0.01             \\ \hline
Conv3 & 4*4*50*500  & 800 * 500     & 0.001            \\ \hline
Conv4 & 1*1*500*10  & 500 * 10      & 0.01             \\ \hline
\end{tabular}
\label{simpleNN_size}
\end{table}
   
   \begin{figure}[t]
      \centering
     % \framebox{\parbox{3in}
      \includegraphics[width=0.7\textwidth]{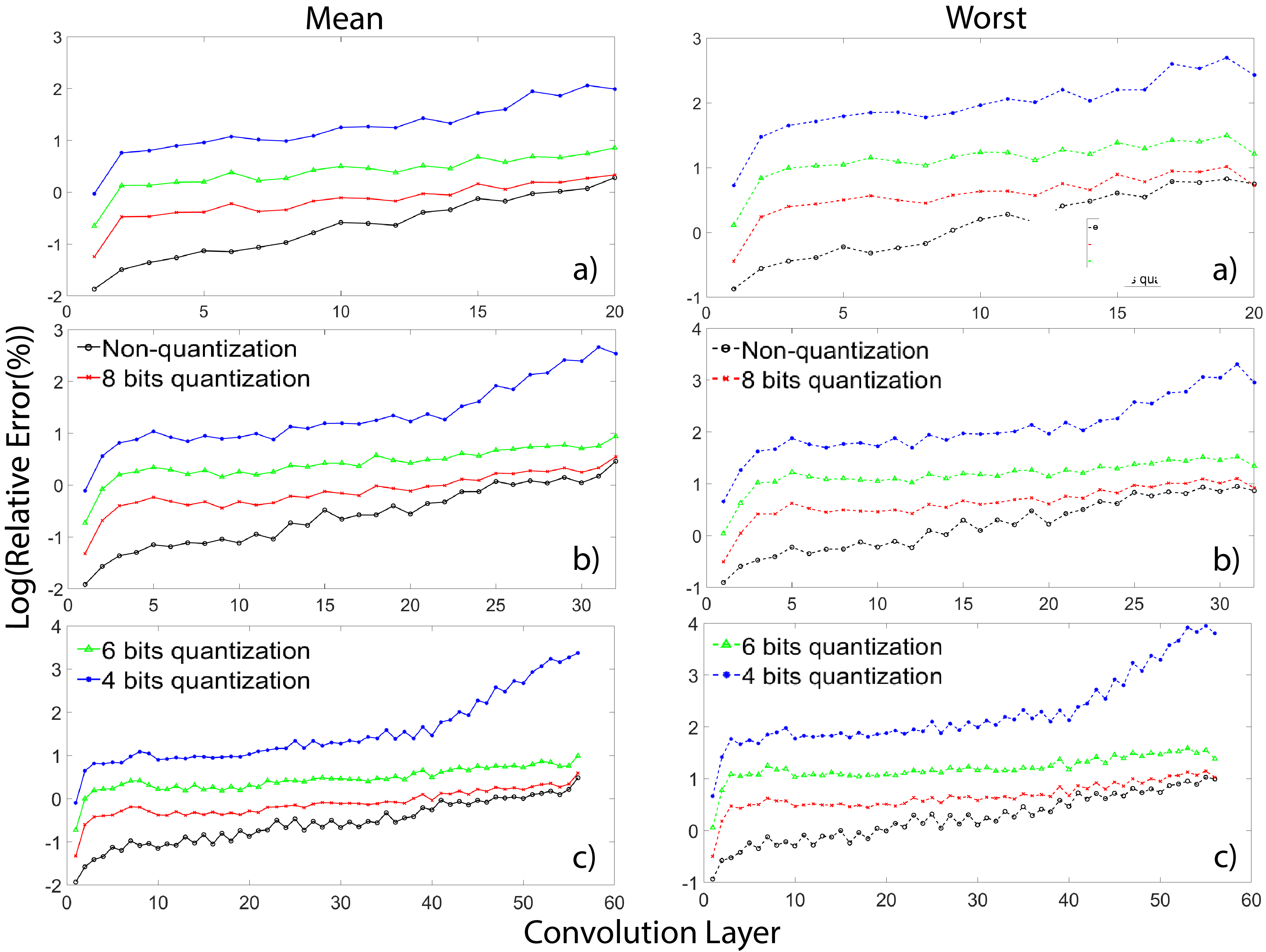}
      \caption{Error propagation along layers in ResNet, solid lines represent mean error, dash lines represent worst error. a) is ResNet-20, b) is ResNet-32, and c) is ResNet-56.}
      \label{fig_layer_error}
   \end{figure}

\begin{table}[]
 \caption{ResNet kernel, crossbar size and conversional signal setting}
\begin{tabular}{|l|l|l|l|l|l|}
\hline
ResNet-20    & ResNet-32    & ResNet-56    & \begin{tabular}[c]{@{}l@{}}Kernel\\ Size\end{tabular} & \begin{tabular}[c]{@{}l@{}}Xbar\\ Size\end{tabular} & \begin{tabular}[c]{@{}l@{}}Conv. \\ Sigal.\end{tabular} \\ \hline
Conv1       & Conv1       & Conv1       & 3*3*3*16                                              & 27*16                                                   & 0.1                                                          \\ \hline
Conv2-7   & Conv2-11  & Conv2-19  & 3*3*16*16                                             & 144*16                                                  & 0.1                                                          \\ \hline
Conv8       & Conv12      & Conv20      & 3*3*16*32                                             & 144*32                                                  & 0.1                                                          \\ \hline
Conv9-13  & Conv13-21 & Conv21-37 & 3*3*32*32                                             & 288*32                                                  & 0.05                                                         \\ \hline
Conv14      & Conv22      & Conv38      & 3*3*32*64                                             & 288*64                                                  & 0.05                                                         \\ \hline
Conv15-19 & Conv23-31 & Conv37-55 & 3*3*64*64                                             & 576*64                                                  & 0.01                                                         \\ \hline
FC          & FC          & FC          & 1*1*64*10                                             & 64*10                                                   & 0.1                                                          \\ \hline
\end{tabular}
\label{resnet_size}
\end{table}
  
   \begin{table}[t]
   \caption{Classification accuracy with different quantization levels}
\begin{tabular}{|l|l|l|l|l|l|}
\hline
Network & Software  & non quantization & 4 bit & 6 bit & 8 bit  \\ \hline
ResNet-20 (CIFAR-10)     & 89.3\%          & 88.7\%         & 11.3\%       & 82.7\%         & 90.7\%             \\ \hline
ResNet-32 (CIFAR-10)     & 89.3\%          & 89.3\%         & N/A         & 82\%          & 90\%             \\ \hline
ResNet-56 (CIFAR-10)     & 90\%           & 88\%          & N/A         & 88\%          & 89.3\%             \\ \hline
LeNet (MNIST)   & 99.1\%  &98.9\%        & 79.2\%       & 88.6\%         & 88.8\%                    \\ \hline
\end{tabular}
\label{more_result}
\end{table}

\subsection{Impact of programming error}

\begin{figure}[t]
      \centering
      \includegraphics[width=0.5\textwidth]{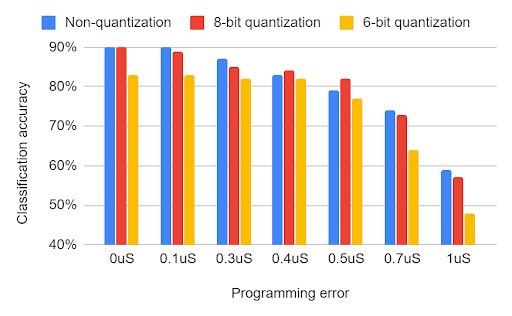}
      \caption{Impact of programming error on classification accuracy. }
      \label{Fig_programming_error}
\end{figure}

We modelled the programming error as conductance variations following zero-mean Gaussian distributions with sigma ranges from 0 to 1 $\mu S$\cite{hu_memristor-based_2018}. 
The devices' conductance ranges from 3.3 $\mu S$ (300k ohm) to 66.7 $\mu S$ (15k ohm).
We would like to emphasize that since crossbar-based computing is based on Ohm's law and KCLs, computing result is carried by current. 
The error current is linearly proportional to the error in conductance, rather than error in resistance. 
Thus, a large resistance variation at high resistance state may not cause large computing error because the relative error in conductance is small. 
Fig. \ref{Fig_programming_error} shows that how programming error affects the final classification accuracy of ResNet-20 on CIFAR-10 with different quantization levels.
To conduct the simulation, we generate the programming error pattern with different sigma settings and add them to the conductance matrices.
Calibration step is then performed on the programming error noise contaminated conductance matrix and update the fitting parameter P.
The above configurations are stored for all input test data to make sure they all use the same crossbar setting. 

When the programming error sigma <0.4$\mu S$, ResNet-20 could still maintain the accuracy higher than 80\%.
Since the 0.4$\mu S$ is less than the 6-bit resolution under our configuration, the 6-bit quantization has minimal impact.
While the non-quantization setting has the maximum accuracy degradation.
As the programming error sigma increase, quantization step could not prevent such error propagating between layers and the classification accuracy drops dramatically.

\section{\bf Conclusions}

In this work, we investigate how modern CNN performs on crossbar-based architecture with end-to-end circuit simulations with careful consideration of parasitic resistances. 
By studying CNN layer by layer, we find CNNs' characteristic like data sparsity cause existed crossbar-based optimization algorithm invalid. 
We propose dense mapping to achieve efficient convolution kernel to crossbar mapping. 
And we adapted and improved the original conversion algorithm for CNNs, which enables 0.25\% mean relative error ($\sim$ 8.6 bits) or 1.2\% worst relative error ($\sim$ 6.4 bits) for crossbar size $576 \times 64$.
We performed a rigorous end-to-end circuit simulation for every convolution layer to give an accurate prediction of error propagation due to analog circuit errors. 
We find that 8-bit or even 6-bit ADC/DAC is necessary to prevent error accumulation in deep CNNs up to 50 layers, and maintains the final classification accuracy. Simulation result also shows that our method is independent of input data sparsity and kernel type. It would be applied to general CNNs to improve their accuracy performance on crossbar-based architecture.

\section{\bf Acknowledgement}
This project is supported by HUAWEI with confirmation number HIRPO2017050311. Any Opinions, findings, and conclusions or recommendations expressed in this material are those of the authors and do not necessarily reflect the views of HUAWEI or its contractors.

\bibliographystyle{ACM-Reference-Format}
\bibliography{DPE}

% \begin{IEEEbiography}[{\includegraphics[width=1in,height=1.25in,clip,keepaspectratio]{./images/Fan_Zhang.JPG}}]%
% {Fan Zhang}
% is currently a PH.D. student in Electrical and Computer Engineering at Binghamton University since 2018 spring. He received his B.S. in Electrical Engineering from Tianjin Chengjian University, China in 2015, M.S. degree in Electrical Engineering from Stevens Institute of Technology in 2017. His research focuses on software/hardware co-design and machine learning algorithm acceleration via emerging device--memristor. 
% \end{IEEEbiography}

% \begin{IEEEbiography}[{\includegraphics[width=1in,height=1.25in,clip,keepaspectratio]{./images/MiaoHu.jpg}}]%
% {Miao Hu} (M'13) is currently an assistant professor at ECE department of Binghamton University since 2017 fall. Before that he worked in Hewlett Pack Labs from 2014 to 2017 on neuromorphic computing as a postdoc. He received his B.S. on Electrical Engineering (Automation) from Shanghai Jiao Tong University, China in 2008, M.S. degree on Electrical Engineering from NYU Tandon School of Engineering in 2011, and Ph.D. degree from University of Pittsburgh in 2014. His research focus on ultra-low power computing via emerging devices such as memristors, and extends to intertwined fields including machine learning software/hardware co-design, lost-cost AI for IoT and artificial nervous systems. He has published more than 40 papers in top conferences and journals, and has more than 30 patents granted or pending.  
% \end{IEEEbiography}

\end{document}